\documentstyle[11pt,paspconf,epsfig]{article}

\begin{document}

\newbox\grsign \setbox\grsign=\hbox{$>$} \newdimen\grdimen \grdimen=\ht\grsign
\newbox\simlessbox \newbox\simgreatbox \newbox\simpropbox
\setbox\simgreatbox=\hbox{\raise.5ex\hbox{$>$}\llap
     {\lower.5ex\hbox{$\sim$}}}\ht1=\grdimen\dp1=0pt
\setbox\simlessbox=\hbox{\raise.5ex\hbox{$<$}\llap
     {\lower.5ex\hbox{$\sim$}}}\ht2=\grdimen\dp2=0pt
\setbox\simpropbox=\hbox{\raise.5ex\hbox{$\propto$}\llap
     {\lower.5ex\hbox{$\sim$}}}\ht2=\grdimen\dp2=0pt
\def\gtrsim{\mathrel{\copy\simgreatbox}}
\def\lesssim{\mathrel{\copy\simlessbox}}
\def\simprop{\mathrel{\copy\simpropbox}}

\def\g{{$\gamma$}}

\title{The Physics of Hybrid Thermal/Non-Thermal Plasmas}

\author{Paolo S. Coppi}
\affil{Astronomy Department, Yale University, P.O. Box 208101
    New Haven, CT 06520-8101, USA}

\begin{abstract}
Models of the continuum radiation from accreting hot plasmas
typically assume that the 
plasma heating mechanism produces energetic particles
distributed in energy either as a Maxwellian 
(the ``thermal'' models) or as an extended power law (the 
``non-thermal'' models). The reality, however, is that
neither description is probably accurate.  In other astrophysical
contexts  where we have been able to
observe the actual particle energy distributions,
e.g. solar system  plasmas, and in
many particle acceleration theories, the heating mechanism
supplies only some fraction of the available energy to 
very energetic particles. The remainder goes into producing
lower energy particles which settle 
into a quasi-Maxwellian energy distribution.
Here, I review the arguments for ``thermal'' versus ``non-thermal''
plasmas in accreting black hole systems
and discuss the physics and emission properties of ``hybrid'' 
plasmas, where the particle distribution energy is approximately
a Maxwellian plus a power law tail. Using results from a new
emission code, I then show that such plasmas
may be relevant to explaining recent observations, particularly
those of Galactic black hole candidates in their soft state.
\end{abstract}



\section{Introduction}

Most attempts at modeling the emission from accreting black hole
systems typically assume that the particle energy distribution
responsible for the emission is either purely ``thermal''
(i.e., a Maxwellian) or purely  ``non-thermal'' (i.e., a 
power law). This is partly due to the desire for 
convenience and simplicity, and partly due to our continued
ignorance as to when and how non-thermal particles exchange energy
and thermalize with neighboring particles. Despite our ignorance,
however, the reality is that Nature probably
never makes particle distributions that are strictly
of one type or the other. As a concrete example which
may be very relevant to the process of black hole accretion, consider
solar flares. (Many analogies have been made between magnetic reconnection
and flare events in the solar corona, and those that might occur
in accretion disk coronae, e.g., Galeev, Rosner, \& Vaiana 1979.)
The WATCH instrument on the GRANAT satellite has recently collected a 
large sample
of flare events observed at energies above $\sim 10$ keV (Crosby et al.
1998). Many of those observations were made in coincidence with
the GOES satellite which measures solar flare at lower 
$\sim$ keV. While the GOES observations usually can be adequately interpreted
in terms of purely thermal emission, in many instances WATCH sees
a significant high-energy excess over the emission predicted
from the GOES spectral model. In other words, many solar flare
events are produced by particle  distributions with 
not only a strong thermal component, but also a significant non-thermal tail:
the emitting plasma is a ``hybrid'' thermal/non-thermal
plasma. If the solar corona-accretion disk corona analogy is correct, 
then something similar may well occur near black holes and could
have important consequences for black hole emission models.

The goals of this contribution are to review the arguments for
and against the existence of similar particle distributions
in the context of accreting black hole systems, to
see what the practical consequences of having a hybrid plasma
might be, and then finally to ask whether we have any concrete
concrete evidence that such plasmas play a role in the
observed emission. In \S 2 below, I examine the theoretical
arguments for and against having a purely thermal 
electron distribution in a black hole accretion disk
corona.  In \S 3, I examine some of the consequences
of having a hybrid plasma and present model spectra calculated
assuming varying amounts of thermal heating and non-thermal acceleration
power are supplied to the plasma. In \S 4, I argue that we may
already have strong evidence for hybrid plasmas in 
the spectra of Galactic Black Hole Candidates, especially
in spectra obtained during their soft state. I show how
a simple phenomenological model based on a hybrid plasma 
can explain the spectra of Cyg X-1 in its soft, hard, 
and transitional states in terms of one basic parameter,
a critical radius outside of which the accretion disk is a
cold Sunyaev-Shakura (1973) disk, and inside of which the disk is a hot,
ADAF-like (Narayan \& Yi 1995) corona.
I summarize my conclusions concerning the importance
and relevance of hybrid plasmas in \S 5.

\section{The Electron Energy Distribution: Maxwellian or Not?}

Consider a relativistic electron 
that finds 
itself in the middle of a low-temperature,
thermal bath of background electrons. Will the electron couple
to the thermal electrons and share its energy with them before
it does something else interesting, e.g., radiate its energy away?
If it does, then it is valid to treat that electron and its energy
as belonging to the thermal pool of electrons.  Before anything
else happens, the electron will have exchanged its energy 
many times with other electrons, which is the requirement 
for setting up a Maxwellian (``thermal'') energy distribution.
Hence, when considering processes that occur on time scales
longer than this ``thermalization'' (energy exchange) time scale,
one is justified in assuming that the electron energy 
distribution at energy $\gamma m_ec^2$ is indeed described by a
relativistic Maxwellian, 
$N(\gamma) \propto \gamma^2\beta\exp(-\gamma m_ec^2/kT_e)$, 
where $\beta$ is the electron velocity, and $T_e$ is the 
temperature of the thermal bath electrons.
If it does not, then
for practical purposes, e.g., when computing the emitted radiation
spectrum,  that electron is decoupled from the thermal distribution.
Assuming that the distribution at this energy is a Maxwellian
could be quite dangerous.
The question of
whether a plasma is best described as thermal or ``hybrid''
(thermal with a significant non-Maxwellian component) thus 
comes down to how rapid the thermalization process is for 
energetic electrons. 

Unfortunately, the details of thermalization
and energy exchange in plasmas are in general poorly understood. 
Witness the current controversy (see the contribution
of Quataert, these proceedings) on whether accretion disks
can support proton and electron energy distributions with
different temperatures (one of the key assumptions behind 
ADAF disk models, e.g., Narayan \& Yi 1995). We know of at least one process
that will exchange energy between protons and electrons
(and similarly between electrons and electrons),
namely Coulomb collisions/scatterings where one particle
is deflected by the electric field of the other particle.
Under the conditions supposed to exist in an ADAF (Advection
Dominated Accretion Flow), the rate
of energy transfer due to Coulomb collision is not rapid
enough and if no other process intervenes, the electrons will
lose most of their energy before sharing it with the protons,
i.e., the protons and electron distributions will not have the
same temperature. But might another energy exchange process be important? 
Possibly. The electromagnetic force is a long-range one,
and plasmas have so-called ``collective modes'' (involving the
coherent interactions of many particles rather than the 
simple interaction between two particles that occurs in a 
Coulomb scattering). Under  certain conditions, these
modes can efficiently couple electrons and/or protons
of different energies even when Coulomb scattering (``two
body relaxation'') is not important.  For an astrophysical example of
such collective effects applied to the proton-electron
coupling problem, see Begelman and Chiueh (1988). (See also Tajima
\& Shibata 1997 and references contained therein for 
a recent overview of collective and other processes in 
astrophysical plasmas.) In solar system plasmas observed
directly by spacecraft, we know that some effect like this
must be operating since  quasi-Maxwellian electron energy distributions
are 
observed even though the Coulomb energy exchange time scales 
are enormous (since the plasma densities are so low).
Similar processes might well be operating in a black hole 
accretion disk or in a corona above the disk.  However, realistic
calculations of collective processes are notoriously difficult
from first principles,  especially near 
an accreting black hole  where 
we still have a relatively poor understanding of the exact physical
conditions. Besides collective processes, we note one other
way for electrons to exchange energy, namely via the photon
field. If the source is optically thick to photons 
emitted by an electron, the absorption 
of those photons by different electrons will effectively transfer 
energy between the electrons and produce a quasi-Maxwellian distribution.
This is the basic principle behind the ``synchrotron boiler'' of
Ghisellini, Guilbert, \& Svensson (1988) (see also
Ghisellini, Haardt, \& Svensson 1998; Svensson, these
proceedings), which can lead
to hot, quasi-thermal electrons distributions. 

If a thermal electron distribution near a black hole can be maintained
only by Coulomb collisions, however, we run into a situation similar to that
of the inadequate electron-proton coupling in ADAFs. In particular,
the photon energy density near an efficiently accreting black
hole is so high that 
energetic electrons cool very rapidly due to Compton scattering. 
To see this,
let us make the standard first-order assumptions 
that the high-energy emission region 
near a black hole is roughly spherical with a characteristic
radius or size $R,$ and that inside this region, the 
photon and particle distributions are uniform and isotropic.
For simplicity, let us also assume that the characteristic energy 
of photons in the source is significantly less than $m_e c^2$ 
($m_e$ is the electron mass, and $c$ is the speed of light) 
so that Klein-Nishina corrections are not important (which
is roughly true for radio quiet AGN and Galactic Black Hole
Candidate systems). In this case, the Compton
cooling rate for an electron of Lorentz factor $\gamma$
is then roughly 
$$\dot \gamma_{comp} 
\approx -{4 \over 3} {\gamma^2 \over m_ec^2}\sigma_T
c U_{rad}$$ where 
$\sigma_T$ is the Thomson cross-section, and $U_{rad}$ is
the characteristic radiation energy density inside the source.
To estimate $U_{rad},$ we note that if 
the source is optically thin, emitted photons will escape the source
on a time scale $\approx R/c,$ the characteristic source light-crossing
time. The total luminosity of the source is then
$$L_{rad} \approx ({\rm Source\ Volume}) \times U_{rad} \times (c/R),$$
which, for a spherical source, gives us 
$$U_{rad} \approx {3 \over 4\pi R^2c}  L_{rad}.$$
A convenient way to work with the source luminosity is to
re-express it in terms of the dimensionless ``compactness''
parameter,
$$l_{rad} \equiv {L_{rad} \over R} {\sigma_T \over m_ec^3}.$$
In terms of this parameter, the cooling rate is then simply
$$\dot \gamma_{Comp} \approx {1\over \pi} \gamma^2 l_{rad} 
\left( \frac{R}{c}\right)^{-1},$$  which gives a characteristic
Compton cooling time
$$t_{cool}(\gamma) \sim {\gamma \over |\dot \gamma_{Comp}|}
 \approx \pi \gamma^{-1} l_{rad}^{-1} \left( \frac{R}{c}\right).$$
Using the total source luminosity and a lower limit
on source size from variability consideration ($R/c \gtrsim \Delta T_{min}$),
many accreting black hole systems are inferred to have compactnesses $l_{rad}$
significantly greater than one (e.g., see Done \& Fabian 1989 for a compilation
of AGN compactnesses; for the Galactic black hole candidate, Cyg X-1, the
best fit compactness seems to be $l_{rad} \sim 10-30$, e.g., see \S 4 below).
In other words, one does not have to go to a very high Lorentz factor 
before $t_{cool} \ll R/c,$ i.e., before Compton cooling becomes very
rapid. Note that if the source is optically thick, then the time 
it takes photons to leave the source is longer, $U_{rad}$ is
correspondingly higher, and the Compton cooling time for
an energetic electron is even shorter. 

To see if this rapid cooling inhibits electron thermalization, 
we must compare this cooling time scale with the corresponding
time scale for an energetic electron to transfer its energy
to a thermal bath of lower energy electrons via Coulomb 
collisions. For purposes of making a rough estimate, we will
assume that the Lorentz factor of the electron is $\gamma \gg 1$
and the thermal electrons are relatively cold ($kT_e \ll m_e c^2$).
Then the  Coulomb ``cooling'' rate of the energetic 
electron is simply given by
(e.g., see discussion and references in \S 5 of Coppi \& Blandford 1990,
also Dermer \& Liang 1989),
$$\dot \gamma_{Coul} \approx -\sigma_T c N_{Th} \ln \Lambda,$$
where $N_{Th}$ is the number density of thermal bath electrons 
and $\ln \Lambda$ is the usual Coulomb logarithm, a weak
function of the plasma parameters with typical value
$\sim 15-30$ for accreting black hole sources. In terms
of the characteristic Thomson (electron scattering) optical depth
of the source, $\tau_T = \sigma_T N_{Th} R,$ this gives us
a Coulomb energy exchange time scale
$$t_{exch}(\gamma) \sim {\gamma \over |\dot \gamma_{Coul}|}
\approx {\gamma \over \tau_T \ln \Lambda} (R/c),$$
which grows with increasing electron energy and eventually becomes
longer than the Compton cooling time (which decreases with 
energy). 
Setting $t_{exch} = t_{cool},$ we obtain the 
Lorentz factor above which electrons cool before thermalizing:
$$\gamma_{th} \approx  \left( \pi \ln \Lambda {\tau_T \over l_{rad}}
\right)^{1/2}.$$
For typical values $\tau_T \sim 1$ and $l_{rad} \sim 20$ and
$\ln \Lambda \sim 20,$ we then have $\gamma_{th} \sim 2,$ i.e., if
Coulomb collisions are the only energy exchange mechanism, thermalization
very quickly becomes ineffective and maintaining a thermal,
Maxwellian energy distribution with relativistic temperature 
$kT_e \gtrsim m_ec^2$ is impossible. (See Ghisellini, Haardt,
\& Fabian 1993 for more discussion along these lines. They
also point that $t_{exch}$ can significantly exceed the variability
time scale $\sim R/c,$ again calling into question validity of 
assuming a single temperature thermal distribution.)
For significant compactnesses
$l_{rad}\sim 20,$ note also that the synchrotron boiler mechanism is 
also quenched by the rapid Compton cooling (Coppi 1992)
unless the source is strongly magnetically dominated,
i.e., the magnetic energy density $U_B \gg U_{rad}$. 
This condition might hold if the particle acceleration
is due to magnetic flares in an active corona, but this is
far from clear yet. 
If, 
as in Cyg X-1, 
we observe photons with energies $\gtrsim m_e c^2$ (511 keV),
one must then seriously consider the possibility that they are 
produced by electrons in the  non-thermal tail of a non-relativistic
quasi-Maxwellian distribution.
One loophole in this
argument is that if a plasma is very photon starved (e.g.,
see Zdziarski, Coppi \& Lightman 1990), i.e., if the power
supplied to the electrons is much larger than the power supplied
to the background photon field, most of $L_{rad}$ will emerge
at energies $\sim m_e c^2.$ In this case, Klein-Nishina corrections
can dramatically decrease the effectiveness of the Compton
cooling and thermalization might be effective to much higher 
energies. The typical AGN and Galactic black hole systems known
to date appear not to be so photon-starved, however. Remember also
that the Compton cooling rate depends critically on the source
luminosity. If the accretion disk radiates very inefficiently
as in the ADAF model proposed for the Galactic center (and the 
black holes at the centers of some elliptical galaxies), 
then $l_{rad}$ can be quite small and thermalization can
be correspondingly effective. For a more detailed discussion
of the electron energy distribution in the context of 
ADAFs, see Mahadevan \& Quataert (1997).
They note that while the
bulk of the electron population may thermalize (especially at
low accretion rates), a significant non-thermal, high-energy
tail is still likely to be present due to synchrotron cooling effects.
Compressional heating effects may also significantly distort the shape of
the electron distribution.

The  preceding argument and the current lack of a
thermalization mechanism that clearly operates 
faster than Coulomb collisions
are not the only reasons to consider a ``hybrid'' electron
energy distribution, where only the lower energy end
is well-described by a Maxwellian.  Observationally, for example,
the electron energy distributions in solar system space plasmas  are never 
exactly described by Maxwellians and often show significant
non-thermal tails. In fact, they sometimes show an energy distribution 
with two (!) quasi-Maxwellian peaks at different energies, i.e., as
if the energy exchange processes are effective only at coupling electrons of
similar energy. Also, whenever we see dissipative 
events, such as solar flares or reconnection events in the Earth's
magnetotail or shocks in the solar wind, we often see direct evidence of
significant particle acceleration accompanied by ``non-thermal" emission.  
This particle acceleration, however, is almost never 100\% efficient 
in the sense that
all the dissipated power ends up in high energy (non-thermal) particles
(e.g., see discussion in Zdziarski, Lightman, \& Maciolek-Niedzwiecki
1993).
Many theoretical particle acceleration models  have an injection 
``problem'' or condition,
depending on one's point of view, where only particles
above some threshold momentum, for example, are accelerated to
very high energies (e.g., see Benka
\& Holman 1994 for a simple model of hard X-ray bursts in 
solar flares, or Blandford \& Eichler 1987 for a discussion
in the context of shock acceleration theories). 
Only above some energy does the relevant
acceleration time scale for a particle become significantly
shorter than its corresponding thermalization/energy exchange 
time scale for that particle. Particles below
this energy are tightly coupled and their acceleration
simply results in bulk heating of the thermal plasma component 
-- i.e., one creates a hot, hybrid plasma.
In accreting black hole systems, 
the problem of determining exactly what electron energy
distribution results from dissipation and acceleration is further
complicated by the fact that radiative cooling can be strong 
and that the dissipation process is highly variable (e.g., the
``steady state'' spectrum of Cyg X-1 that one fits may
really be the superposition of many shot-like events with
time-varying spectra, e.g.,
as explicitly illustrated by Poutanen \& Fabian in these proceedings). 
The most heroic
effort to date in terms of incorporating a
particle acceleration
scenario into a time-dependent code and then actually 
computing a spectrum is probably that of Li, Kusunose,
\& Liang (1996). See the review by Li (these proceedings)
for some of the more recent results obtained with this
code as well as a detailed discussion of the particle
acceleration problem. The physics incorporated into this
code may ultimately turn out to be overly simple or
simply not relevant (given our large ignorance in this 
problem), but the code nonetheless provides an 
explicit demonstration of how a ``hybrid'' energy distribution might arise.

One final complication is that the dissipation rate and non-thermal
acceleration efficiency may vary considerably inside the 
source (as observations by the Yokoh satellite have shown 
to indeed be the case for solar flares). If the source is only
moderately optically thin ($\tau_T \sim 1$), escaping photons
can traverse the source and sample several regions
with different acceleration efficiencies.
The effective
electron distribution required to produce the observed 
emission will then be a spatial average over the source
and may not be predictable by any~(!) standard one-zone particle acceleration
calculation.  (The same is true if one averages the 
emission over time -- the effective electron distribution
is then some time average and depends critically on the variability
properties of the dissipation regions.)
The bottom line is that spectral modelers should not be
surprised to find that their data requires a somewhat ``strange''  
electron energy distribution.
Given the increasing quality of data (uncertainties in 
2-10 keV X-ray spectra are now down to the few percent level and are 
often dominated by systematics not statistics), it is frankly
amazing that the simple models employed to date have worked as well as
they have. (In the next section, though, I will show why this may
not be quite as surprising as one might first think.)

\section{The Consequences of a ``Hybrid'' Electron Distribution}

Hybrid emission models have been considered for some time in the
context of solar flares, e.g., see  Benka \& Holman (1994), and 
more recently Benz \& Krucker (1999).
As the relevant processes and parameters in these plasmas
are quite different
from those near accreting black holes (e.g., the ambient
radiation energy density is much higher in the black hole case),
I will not consider these models further here. 
To make the discussion more concrete, 
I will show results from one specific incarnation of a plasma
model intended for black hole applications, namely a plasma
code called {\sc eqpair} (Coppi, Madejski, \& Zdziarski 1999)
that we have recently developed. The code is an extension
of the code of Coppi (1992), which makes no intrinsic assumptions
about the electron spectrum other than the lowest energy
electrons are in a quasi-Maxwellian distribution. 

In terms of the basic physics considered, the code is essentially
the same as other codes that have been recently applied to
study hybrid plasma emission: Zdziarski, Lightman, \& Maciolek-Niedzwiecki
(1993), Ghisellini, Haardt \& Fabian (1993), Li, Kusunose, \& Liang (1996),
 and Ghisellini, Haardt, \& Svensson (1998). The source
geometry assumed in these models is either  a disk-corona (slab)
geometry (Ghisellini, Haardt \& Svensson model),
 or a spherical geometry (other models). Low energy 
(UV or X-ray) thermal photons from the accretion disk  are assumed
to be emitted uniformly inside the source region for the spherical
models and enter from the base of the corona in the disk-corona model.
The total luminosity of these ``soft'' photons is parameterized by
a compactness parameter $l_s$ analogous to $l_{rad}$ (although remember
that $l_{rad}$ is defined in terms of the {\it total} photon luminosity
escaping the source). The spectrum of these soft photons is typically
assumed to be a blackbody or disk blackbody (Mitsuda 
et al. 1984) with characteristic temperature $T_{bb}.$
The main practical difference of the two geometry assumptions is that spherical
models do not show the strong ``anisotropy  break'' (e.g.,
Stern et al. 1995) in the 2-10 keV region that the disk-corona models do. 
To date, strong evidence for such a break has not been seen.

In the Li et al. model, a turbulent wave spectrum is then specified 
and electrons from a cool background thermal plasma with Thomson optical
depth $\tau_p$ are  accelerated by the
waves to form the non-thermal tail. The total luminosity supplied to the
waves and eventually the electrons is specified by the compactness
parameter $l_h$ ($L_h \sigma_T/R m_e c^3$ where $L_h$ is the wave
luminosity).
In the other models, the 
exact acceleration and heating mechanism for the electrons is left 
unspecified. Rather, electrons and/or positrons are injected into the
source with some fixed (non-thermal) energy spectrum $Q(\gamma)$ that does not
need depend on conditions in the plasma. The typical assumptions are
that $Q(\gamma)$ is either a power law extending from Lorentz factor
$\gamma_{min}\sim 1-2$ to Lorentz factor $\gamma_{max} \sim 3-1000,$
a delta function at Lorentz factor $\gamma_{inj} \sim 3-1000,$ or 
an exponentially truncated power law at some energy $\gamma_c \sim
3-1000.$ The total luminosity supplied to the source via these
injected non-thermal pairs is parameterized by a compactness
parameter $l_{nth}.$ To mimic, for example, an acceleration 
mechanism with 
less than 100\% non-thermal efficiency, additional power, parametrized
by the compactness parameter $l_{th},$ is supplied to 
the electrons (or pairs) once they have thermalized.
Background electrons which participate in the thermal
pool may be present, and in {\sc eqpair}, pair balance is not automatically
assumed.
(In electron-positron pair plasmas, pair balance means that the total pair
annihilation rate in the plasma equals the pair creation
rate and is the standard condition used 
to set the equilibrium density in the source, e.g., see
the discussion in Svensson 1984.)
Hot electrons and/or positrons are
allowed to cool via Compton scattering, synchrotron
radiation (so far included only in the Ghisellini 
et al. 1998 code, which, however, does not include pair
processes), 
Coulomb energy exchange with colder thermal pairs, and
bremsstrahlung emission. The energetic photons created by these
processes either escape the source, produce pairs on lower
energy photons, Compton scatter off more electrons, or
are synchrotron self-absorbed.  
Pairs that are injected or created in the plasma
annihilate away once they have cooled. 
Any ``excess'' electrons that were injected
(without a positron partner)
are assumed to be removed from the system once they
thermalize, e.g., via reacceleration, so that no net particles
are added to the system except via pair creation. In steady state,
the escaping photon luminosity must equal the sum of the
various input luminosities, i.e., $l_{rad} = l_{h} + l_{s}$
where for the codes with no acceleration prescription, $l_{h} = l_{th}
+ l_{nth}.$ 

To summarize then, the main parameters of a  hybrid model like
{\sc eqpair} are: (i)
$l_h=l_{nth}+l_{h},$ (ii) $l_s,$ (iii) $l_{nth}/l_{th},$ (iv)
the source radius $R$ (which enters into some of the rate coefficients and 
relates the compactness parameters to absolute source luminosities),
(v) $\tau_p,$ the optical depth of the background electron-proton
plasma, (vi) the characteristic soft photon energy as specified by
$T_{bb},$ and (vii) the non-thermal electron/pair injection spectrum 
$Q(\gamma).$ The usual spectral modeling parameters of 
Comptonization models, e.g.,
temperature of the electron distribution, $T_e,$ and the total Thomson
scattering optical $\tau_T,$ are all computed self-consistently.
The main advantages of {\sc eqpair} over other the codes  is that: (i) all the 
microphysics is treated self-consistently without significant 
approximations (in particular all Klein-Nishina corrections are included,
which is crucial), and (ii) it is still fast
enough to use for real data fitting (one model iteration takes 
$\sim 5-15$ sec on a 300 MHz Pentium II). The code has been ported
to XSPEC and incorporates ionized Compton reflection (as in the 
{\sc pexriv} XSPEC model) including smearing due to relativistic
motion in the disk (as in the {\sc diskline} XSPEC model).  
(Note that in the current version,
the reflected radiation is assumed {\it not} to 
pass back through the emission region as it 
in fact might in a slab-like disk-corona geometry such as that
of Haardt 1993.)
The code
will be available for general use once the description/user's manual
paper is (finally) submitted. One drawback of {\sc eqpair} and the
other codes (except for Ghisellini et al. 1998 who
calculate the Compton upscattered spectrum from the
corona by solving the radiative transfer equation in a slab 
geometry)
is the use of an escape probability to handle the radiation 
transfer. This is a potentially serious limitation when 
the optical depth $\tau_T \sim 1$ and the electron temperature
is high, $\gtrsim 100$ keV. However, as noted in Coppi (1992),
the errors are typically less than the uncertainties 
introduced by our ignorance of the exact source
geometry, e.g., are the 
soft photons actually emitted in the center of the corona
or do they enter the corona from some outer, cool region
of the disk (as in Poutanen, Krolik, \& Ryde 1997)?
Figure~7 below shows
an explicit comparison of the {\sc eqpair} output
versus the output from a Monte Carlo simulation with
three different soft photon injection scenarios.
The optical
depth in that calculation is low ($\tau_T =0.1$) but the temperature is 
rather high ($kT_e=200$ keV), and the agreement is quite reasonable.
One final caveat on using {\sc eqpair} to fit observed spectra is that
the spectrum it produces is of course a steady-state one, while
the real spectra that are being fit are
typically time integrations over many flares.

To illustrate the effects of a hybrid electron distribution 
(or, equivalently, simultaneously accelerating 
non-thermal particles and heating thermal
particles), let us first consider the transition from a 
purely ``non-thermal'' plasma ($l_{th}=0$) to a mainly thermal one with
$l_{th} \gg l_{nth}$. This is shown in Figure~1 where all plasma
parameters are kept fixed except $l_{th}.$ The initial plasma configuration
has $l_{rad}=l_s+l_h=20,$ which means that Compton cooling of pairs is strong
and pair production of gamma-rays on X-rays is moderately 
important. (The compactness parameter not only measures the effectiveness
of Compton cooling, but also the optical depth to photon-photon pair
production in the source, e.g., see Guilbert, Fabian \& Rees 1983.)
Because the non-thermal electrons were ``injected'' with a fixed
Lorentz factor $\gamma_{inj} = 10^3,$ the cooled electron distribution
should have been $N(\gamma) \propto \gamma^{-2}$ which should have 
given a power law photon energy distribution with $F_E \propto E^{-0.5}$. The
deviations from this power law are the result of photon-photon
pair production which removes some of the highest energy photons
and adds lower energy pairs to the electron distribution (see,
e.g., Svensson 1987 for a discussion of how pair ``cascading'' transforms
spectra). Also visible is a pair annihilation feature at 
$\sim$ 511 keV  caused by the 
annihilation of cooled, essentially thermal pairs. Because Compton 
cooling is so rapid, the pairs do not annihilate or thermalize 
until they have already lost most of their energy. (The temperature
of the cool, thermalized pairs is only $\approx$ 10 keV.)
Hence, the annihilation feature is narrow and has the shape expected
from the annihilation of pairs with a thermal distribution. 

Note 
an interesting effect. The initially non-thermal plasma has, in fact,
already turned itself into a hybrid thermal/non-thermal plasma.
Depending on the exact plasma parameters, Compton upscattering of
the soft photons by the thermal component can be quite
important and will produce a soft X-ray excess (see Zdziarski
\& Coppi 1991) without having to invoke any additional emission
component. In Figure~1, we see  this soft excess emerging and 
becoming increasingly visible
as we increase $l_{th}$ and make the cooled thermalized
pairs hotter. Note that as the soft X-ray excess increases, the 
pair production optical depth for gamma-rays increases
correspondingly,  and the flux above 511 keV drops. Eventually as
we keep increasing $l_{th},$ by about $l_{th} = 50$ or
$l_{th}/l_{nth}=5,$ the soft ``excess'' is no longer really an excess but 
in fact dominates the entire X-ray spectrum. The spectrum then is essentially
that of a thermal plasma except for a pronounced gamma-ray excess and
a hint of a broad annihilation line above $\sim$ 100 keV. 
Note the behavior of the annihilation line.
Even though the importance of pair production continually increases
with $l_{th}$ and the annihilation flux is actually always growing, the
annihilation feature eventually disappears as it is broadened and
downscattered (e.g., see  Maciolek-Niedzwiecki, Zdziarski, \& Coppi 1995).
By $l_{th} = 300,$ the spectrum is very close to that
expected from a purely thermal plasma  --
although a detector with sufficient sensitivity and
energy resolution above $\sim 500$ keV would still find an excess
relative to the purely thermal model since the hybrid gamma-ray does
{\it not} fall off exponentially. This gamma-ray excess will be the 
key in understanding the effects of having a hybrid plasma in pair balance
(see below). In sum,
as Figure~1 shows, the most unambiguous signature of a 
hybrid plasma would be the presence of excess emission above $\gtrsim
200$ keV. Unfortunately, if the hybrid plasma is
thermally dominated ($l_{th}/l_{nth} \gg 1$), hot, and moderately
optically thick ($\tau_T \approx 2$ and $T_e \approx 75$ keV for 
the $l_{th}=300$ model), it becomes very hard to distinguish 
spectroscopically from a 
purely thermal one 
-- except at the highest energies $\gtrsim 500$ keV. Any firm conclusions,
for example, on the nature of the plasma in Cyg X-1's hard state
(which is likely to be thermally dominated, hot, and moderately
optically thick)
will have to await better detectors in this energy range like Astro-E
and INTEGRAL, although we note that BATSE and COMPTEL may already
have detected such an excess (see McConnell et al. 1994,
and Ling et al. 1997).

\begin{figure}[t]
\centerline{\epsfig{file=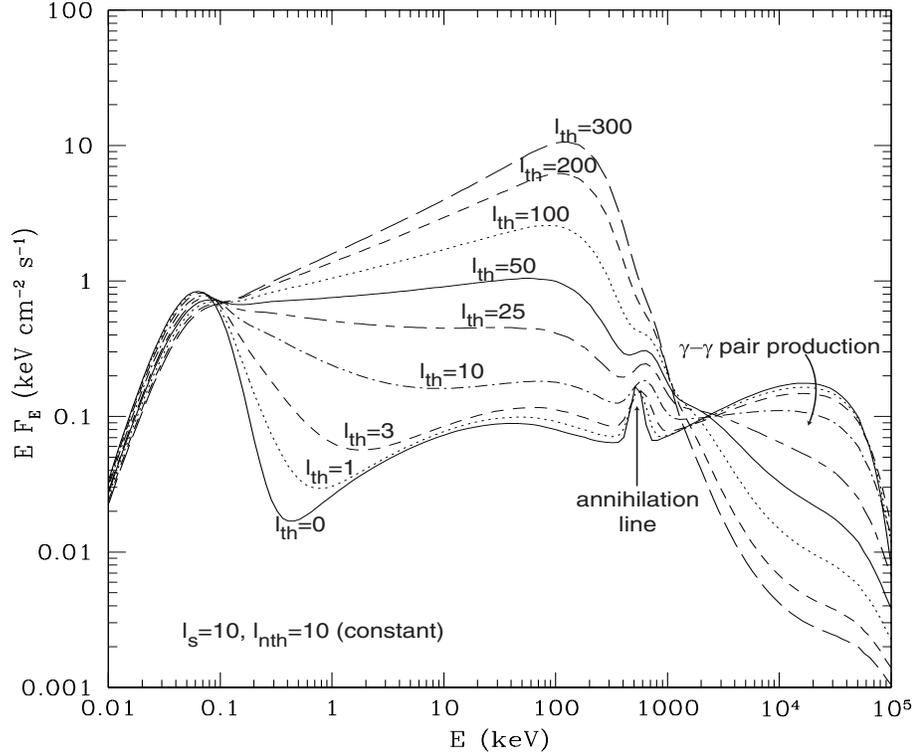,width=12cm,height=10cm}}
\caption{The transition from a non-thermal plasma ($l_{th}=0$)
to a thermally dominated plasma ($l_{th}/l_{nth} = 30$). The soft
input into the source has a compactness $l_s=10$ and has
a blackbody spectrum with $T_{bb} = 15$ eV. The assumed source
radius is $R=10^{14}$ cm, and a  background plasma is  present
with optical depth $\tau_p = 0.1.$ }
\label{fig-1}
\end{figure}

The question of how much the details of the electron energy
distribution matter is unfortunately somewhat more complicated
than the preceding example would suggest. 
The real test seems
to be whether or not multiple Compton scattering is important.
In a typical thermal model with $kT_e \ll m_e c^2,$ the fractional
energy shift per scattering is $\Delta \epsilon/ \epsilon
\approx 4kT_e/m_ec^2 \ll 1.$ In order to cool the thermal
electrons and boost the input
photons to sufficiently high energies that the escaping photon
luminosity equals the total input luminosity, a photon must
scatter many times off the  electrons before it
escapes. This is particularly true if $l_h/l_s \gg 1$
and a significant energy boost is required to satisfy
the energy conservation requirement $l_{rad} = l_h+l_s$. The final emergent
spectrum in a thermal model is thus typically composed of many
so-called ``orders'' of Compton scattering. (Each order of
Compton scattering is calculated by computing the Compton 
scattered photon assuming the previous order as the input 
photon spectrum. The initial soft photon input spectrum is
the zeroth order.)  Since Compton scattering by hot electrons
smears out spectral features (an input photon with a given initial energy
can be scattered to a range of final energies), the spectrum
of each successive Compton scattering order tends to appear 
smoother. The end result is that any spectral features
in the first order of Compton scattering (e.g., due
to the choice of electron energy distribution) tend to be washed out
and the composite emergent spectrum is usually a rather featureless
power law. As shown, e.g., in Rybicki \& Lightman (1979), the
slope of this power law can be derived by basically knowing only the
mean photon energy change per scattering and the mean number of 
scatterings a photon undergoes before escaping (i.e., 
the Compton $y$ parameter).  If one replaces the thermal
electron distribution by another one that gives the 
same mean photon energy shift per scattering and also insures
that the Thomson optical depth of the source remains constant,
then to first order, nothing changes in the preceding chain
of reasoning and the emergent spectrum will be the same(!).
This was noted by Zdziarski, Coppi, \& Lightman (1990) in the
context of photon-starved plasmas and plasmas
with very steep non-thermal injection 
($Q(\gamma) \propto \gamma^{-\Gamma}$ with $\Gamma\gtrsim 3$)
extending to a $\gamma_{min}$ close to unity, and by 
Ghisellini, Haardt, \& Fabian (1993) who showed that
the non-thermal Comptonization spectra produced in plasmas
where $Q(\gamma)$ goes to zero for  $\gamma > \gamma_{max} \sim 2-4$
are very close  to thermal ones where the mean energy per
scattering is the same.  In other words, {\it as long as most of 
the electrons in the source are low energy and multiple
orders of Compton scattering are important, it makes little difference
what energy distribution the electrons have.} 
If non-thermal
electron acceleration  near the black hole 
holes is not very effective, i.e., if electrons never reach
very high energies (perhaps because the radiative cooling times
are so short), this might help explain why objects like Cyg X-1
have thermal-looking spectra in their hard state.
It also explains why different hybrid plasma codes can use rather  
different criteria
for deciding when exactly an electron has thermalized and still 
end up predicting similar emergent spectra.

\begin{figure}[t]
\centerline{\epsfig{file=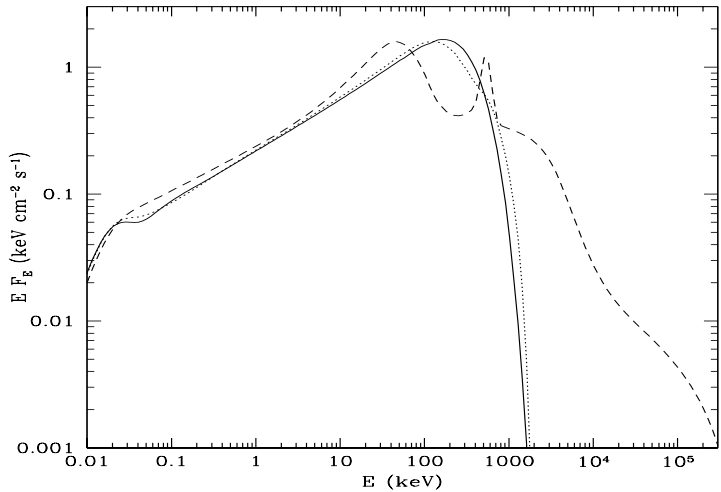,width=10cm,height=7cm}}
\caption{Comparison of spectra from photon-starved thermal and hybrid
models. The {\it solid} curve is the emergent spectrum from a purely
thermal plasma ($l_{nth}=0$) with input parameters $l_s = 3$
and $l_h = 160.$ (The other model parameters are $R=10^{15}$ cm,
$\tau_p = 0,$ and blackbody soft photon injection with
$T_{bb} = 5$ eV.) The plasma temperature and optical depth
derived for this set of parameters is $T_e = 114$ keV and
$\tau_T = 1.2$. The {\it dotted} curve shows the spectrum obtained
for the same model parameters except that now $l_{nth}/l_{th}=4$
and $Q(\gamma) \propto \delta(\gamma-3.6),$ i.e., the model
is a hybrid plasma with low energy electron injection.  The 
{\it dashed} curve shows the spectrum obtained using the same model parameters
as the dotted curve, except that now $Q(\gamma) \propto \delta(\gamma-1000),$
i.e., the model is a hybrid plasma with high energy injection. }
\label{fig-2}
\end{figure}

As a further illustration of how spectra from
different electron distributions can be quite similar
if multiple scattering is important, we show in Figure~2 the spectra
produced by a strictly thermal plasma, by a hybrid plasma where
the non-thermal electron injection function is a delta function
at $\gamma_{inj}  = 3.6,$ and by a hybrid plasma where 
$\gamma_{inj} = 1000.$ For all three plasmas, 
$l_h/l_s \approx 50,$ i.e., the plasma is photon-starved
and multiple scattering is important. The spectra 
from the  thermal plasma and the hybrid, low $\gamma_{inj}$ plasma
are rather similar (particularly in the 1-100 keV 
range), even though no effort was made to tune
the non-thermal electron injection (e.g., as in Ghisellini, 
Haardt, \& Fabian 1993) to match the mean photon energy change
with that of the thermal plasma. The first key reason for
this is that the
equilibrium plasma parameters are determined by pair balance
($\tau_p = 0$) and the fact that the
electron-positron distribution responsible for the
multiple Compton scattering is dominated by pairs created in the source,  
not the injected ones. If the input electron spectra are at all similar
(i.e., have similar mean energies as is the case here), the 
different pair cascade generations  initiated by these electrons
tend to converge -- leading to similar final electron
distributions. The second reason is that when multiple scattering
is important, the upscattered spectrum must pivot about/start from the peak
energy of the injected soft photon spectrum and then will turn down
once the photon energy is comparable to the
 maximum average energy of the scattering electrons.
If the maximum energies of the pairs created in the source
 are at all similar, then simple energy conservation guarantees that the slopes
of the Compton upscattered spectra will be correspondingly similar.
For these reasons, the spectrum obtained in the hybrid plasma
model with high energy non-thermal electron
injection turns out to be amazingly similar (given
the radically different injection spectrum) to the other two spectra.
The agreement between cases only increases as the plasmas
becomes more photon-starved.  

While the overall spectra from hybrid plasmas can be quite 
similar to those from thermal plasmas,  hybrid plasmas with
pairs are also systematically different from thermal ones in that
the larger the excess emission they have at gamma-ray energies
 (i.e., the higher
the typical injection energy of the non-thermal electrons), the
{\it lower} the characteristic equilibrium temperature of the 
cooled pairs in the source. (Note the clearly separated Comptonization and 
annihilation peaks in the dashed spectrum of Fig. 2.) This is because
hybrid plasmas with an energetically insignificant high
energy electron tail will still produce many more pairs for a given thermalized
pair temperature than will a purely thermal plasma. The
gamma-ray spectrum in a thermal plasma is a Wien spectrum and 
cuts off exponentially at photon energies $\gtrsim kT_e.$ As noted
above, however, the gamma-ray spectrum in a hybrid model may fall off 
much more slowly with energy. This means that while the ``pair thermostat'' of 
Svensson (1984) still operates in hybrid models (in general, for a given  
$l_h/l_s,$  the higher $l_h$ the higher the  density of thermalized
pairs and the lower their temperature), the exact plasma parameters
it predicts depend critically on $l_{nth}/l_{th}$ and the
non-thermal injection spectrum.

\begin{figure}[p]
\centerline{\epsfig{file=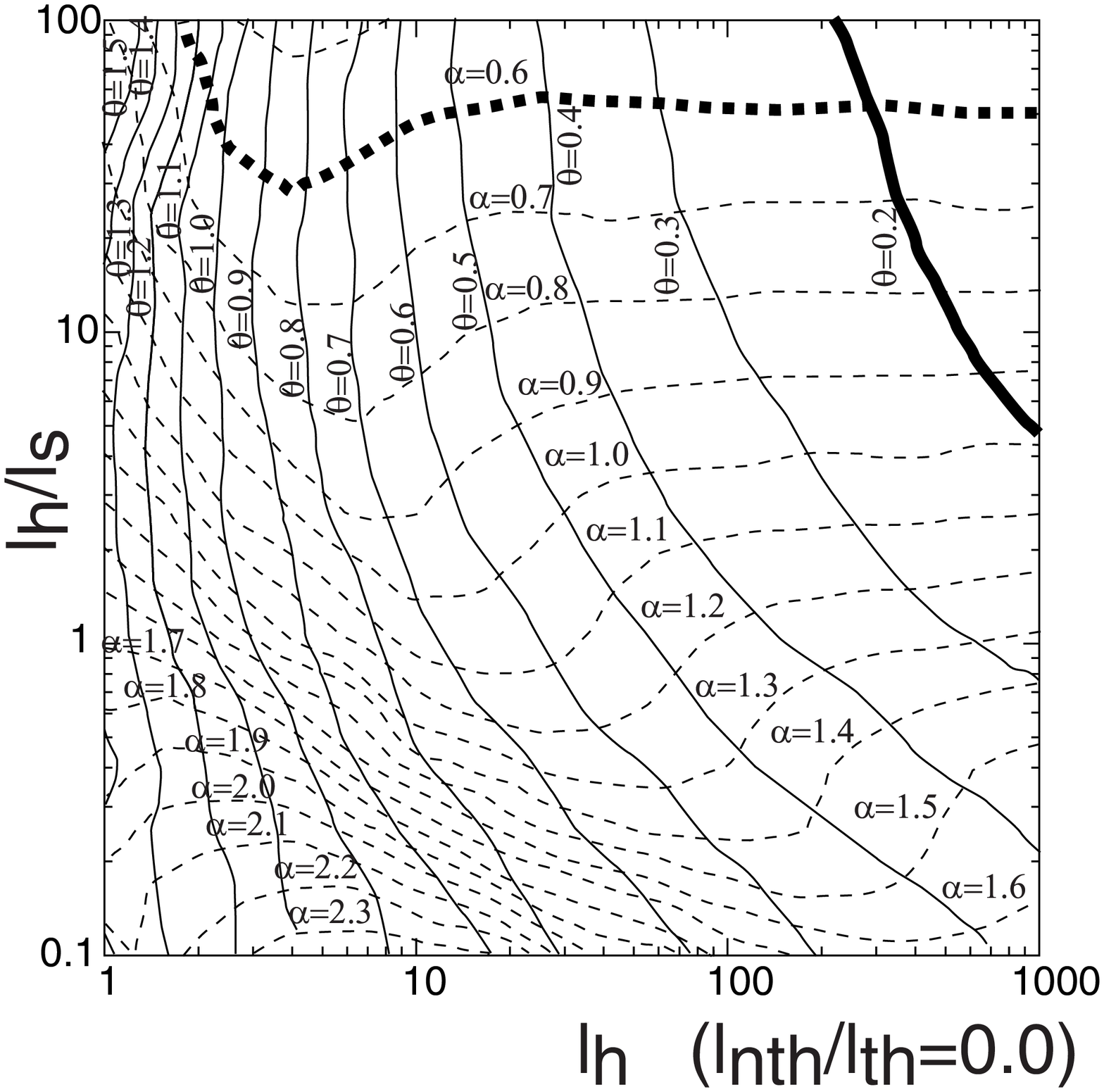,width=10cm,height=7.5cm}} 
\centerline{\epsfig{file=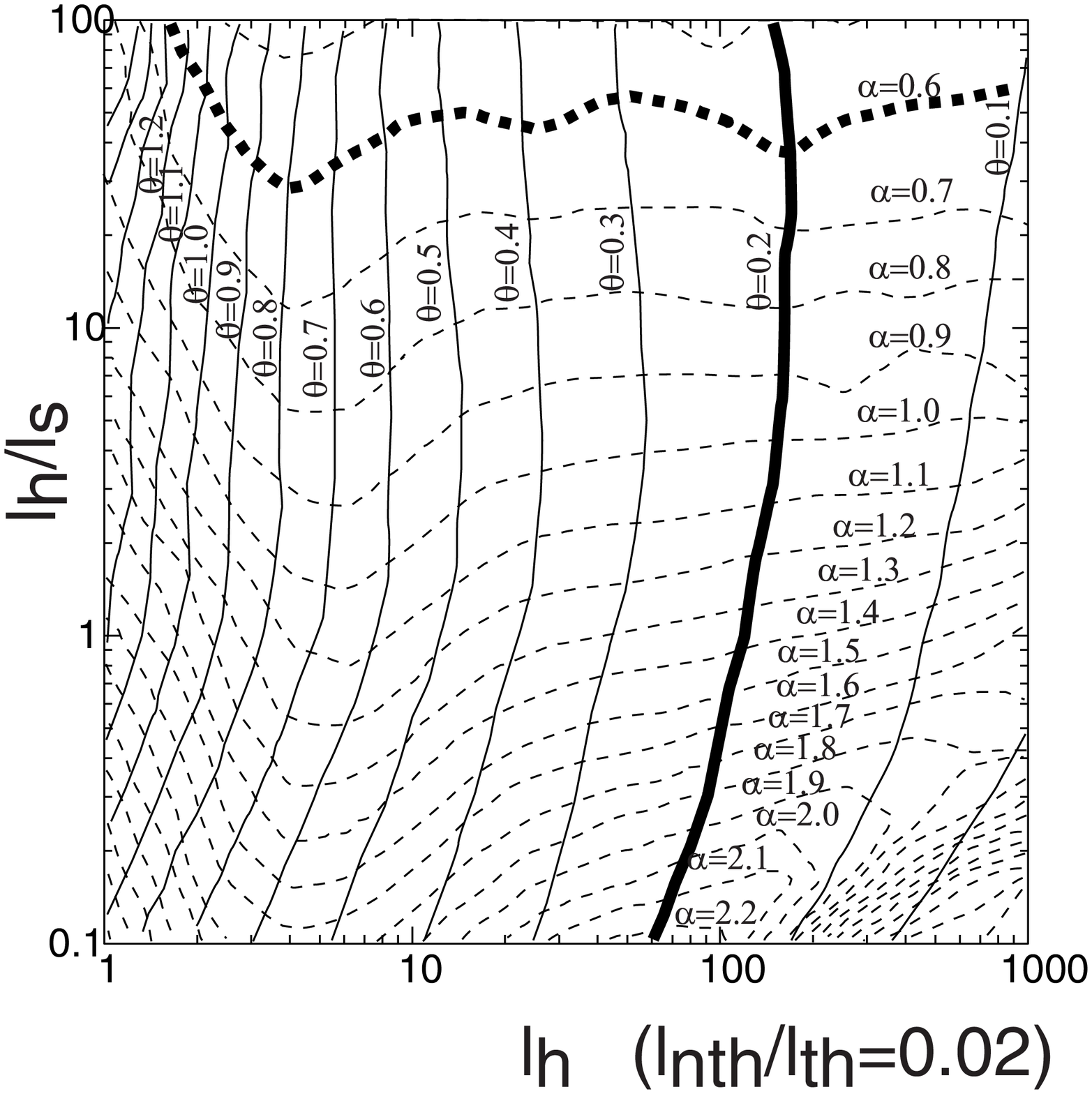,width=10cm,height=7.5cm}} 
\caption{Contours of constant $\alpha_x,$ the 2-10 keV spectral 
index ({\it dashed} curves), and constant $\theta_e=kT_e/m_ec^2,$
the dimensionless plasma temperature ({\it dotted} curves),
in the plane $l_h/l_s$ vs. $l_h$ (see text). The {\it heavy dashed}
curves represent contours with $\alpha_x = 0.6,$ while the {\it heavy
solid} curves are contours with $\theta_e=0.2.$ The 
upper panel (Fig. 3a) shows the results for the pure thermal case ($l_{nth}=0$).
The lower panel (Fig. 3b) shows results for $l_{nth}/l_{th}=0.02,$
and the panel on the next page (Fig. 3c) for $l_{nth}/l_{th}=4.$ 
The {\it vertical} labels give the value of $\theta_e$ for the adjacent 
solid curve, and the {\it horizontal} labels give the value of $\alpha_x$
for the dashed curves. The input energy distribution for the
soft photons was a blackbody with temperature $kT_{bb}$ = 10 eV.}
\label{fig-3}
\end{figure}

\setcounter{figure}{2}
\begin{figure}[t]
\centerline{\epsfig{file=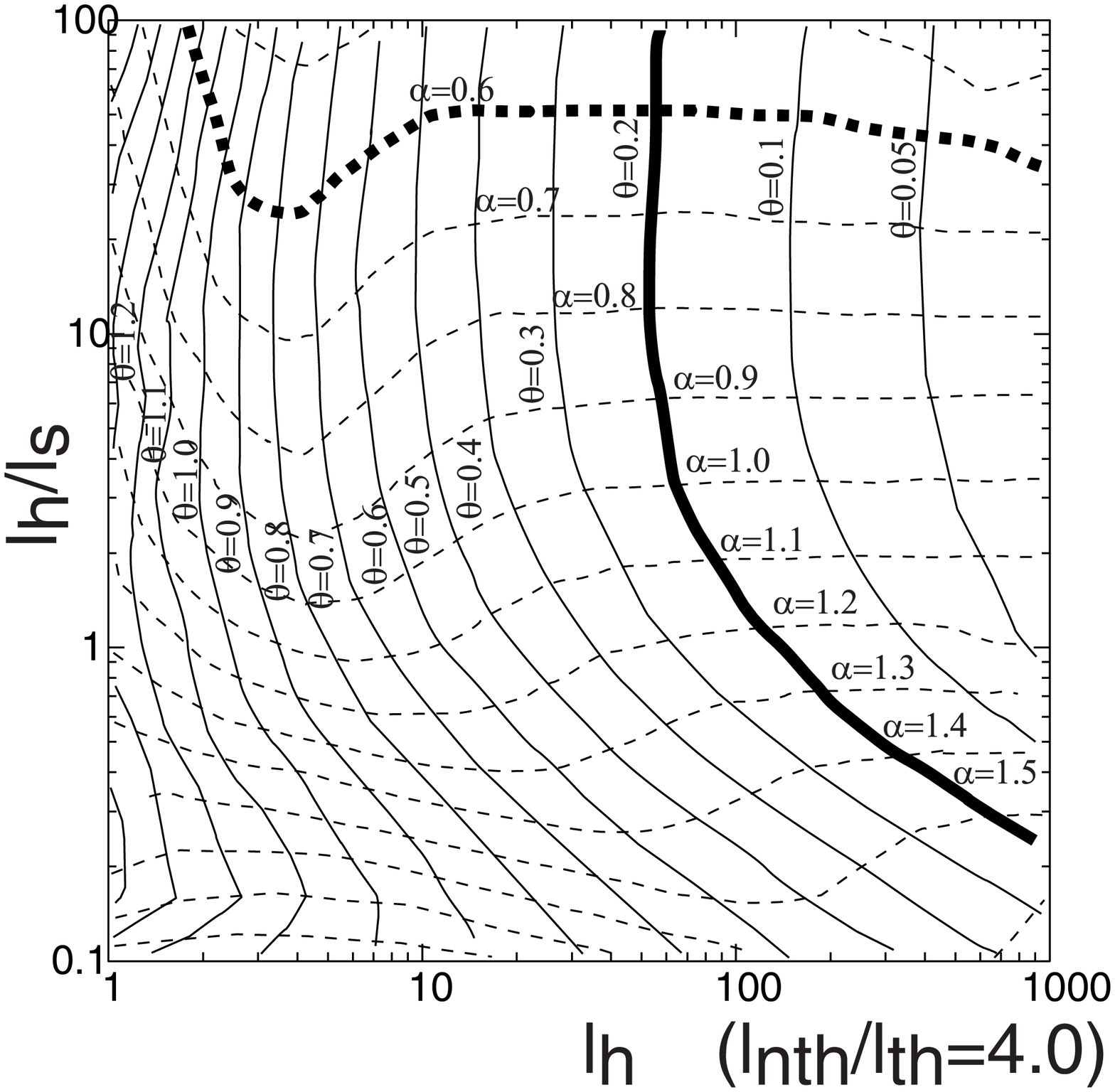,width=10cm,height=7.5cm}} 
\caption{(continued) This panel (Fig.~3c) shows results  
for $l_{nth}/l_{th}=4$.}
\end{figure}

To demonstrate this last point, Figure~3 shows three plots analogous
to fig. 2 of Ghisellini \& Haardt (1994). Figure~3a represents exactly the 
same case as their fig. 2 and shows contours of constant 2--10 keV spectral
index and plasma temperature plotted in the plane of $l_h/l_s$ versus
$l_h$ for a purely thermal model (with no background plasma present).
The two figures agree well given the differences in the microphysics
({\sc eqpair} includes electron-electron bremsstrahlung cooling which is  
important at low compactnesses.)
Figure~3b shows similar contours, but now the plasma model has a small
non-thermal, high-energy ($\gamma_{inj}
=1000$) component that receives only $2\%$ of the total power
provided to the electrons and pairs. While the contours look rather
similar in the photon-starved, high compactness region ($l_h/l_s\gg1,$
$\l_h \gg 1$) for the reasons explained above, 
they behave rather differently in the rest  of the 
diagram. Figure~3c shows what happens when $l_{nth}/l_{th}=4$ 
(the plasma is mainly non-thermal) but the injected electrons
are low energy with $\gamma_{inj}=3.6$ (the case in our Fig.~2).
From the shape of the contours, we see that such a model
indeed behaves much more like
a purely thermal model. However, there are still considerable
differences, e.g., in the predicted spectral index, for $l_h/l_s,
l_h \lesssim 10.$ The lesson here is that while there always seems
to be a rough one-to-one mapping between $l_h/l_s$
and the observed spectral index (higher $l_h/l_s$ gives
harder spectra) and between $l_h$ and the thermal
electron/pair temperature (higher $l_h$ gives lower temperatures),
the details of the mapping are {\it not} robust. 
If one removes the constraint of pair balance by
allowing a non-zero $\tau_p,$ the mapping changes even more.
When the time comes to extract the physical plasma parameters from
the observed spectra, the possibly hybrid nature of the
plasma energy distribution can make a significant difference -- even 
if the observed $1-200$ keV spectrum appears consistent
with pure thermal Comptonization. This having been 
said, figures of the type shown in Figure~3 and Ghisellini \& 
Haardt (1994) should still be very useful. Given a particular set
of assumptions about the plasma, they allow one to immediately
zero in on the relevant model parameter space. In this regard,
we also direct the reader to the contribution of Beloborodov
(these proceedings) where some new analytic approximations
for thermal Comptonization are presented.

\begin{figure}[t]
\centerline{\epsfig{file=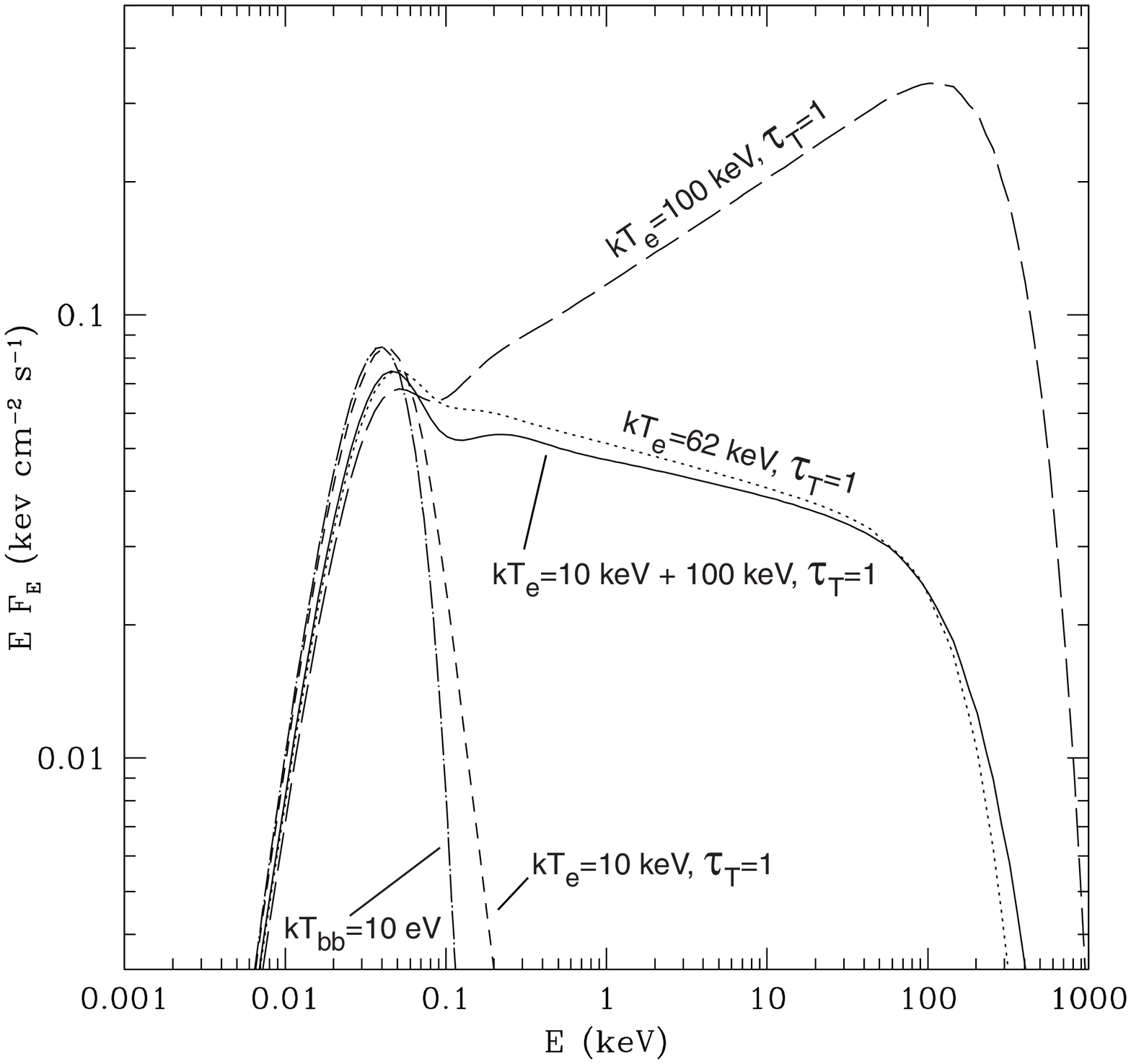,width=10cm,height=8cm}}
\caption{Comparison of Comptonization spectra produced by 
thermal electron energy distributions
({\it dotted}, {\it dashed},  {\it long-dashed} curves)
versus the spectrum from a hybrid distribution ({\it solid} curve) 
which is the sum of two Maxwellians of different temperature (see text).  
The {\it dashed-dotted} curve
shows the input (unscattered) blackbody soft photon spectrum.
The parameter $\tau_T=1$ is the Thomson optical depth of the 
spherical source region. Spectra are computed using the 
Comptonization routine of {\sc eqpair}.}
\label{fig-4}
\end{figure}

We conclude this section by showing one final example 
of how only the mean fractional change in photon energy per scattering 
matters when multiple Compton scattering occurs. In Figure~4, we
set the electron distribution equal to the sum of two Maxwellians,
one with $T_e^{lo}=10$ keV and one with $T_e^{hi}=100$ keV. This might at
first seem a silly choice, but it is not. Several authors 
(e.g., Liang 1991 and Moskalenko, Collman, and Sch\"onfelder
1998) have attempted to explain the possible hard tail in Cyg X-1
via a multi-zone model where thermal electrons have significantly
different temperatures. The idea is that Compton upscattering
in the lower temperature zones explains the X-ray emission,
while upscattering in the higher temperature zones explains the
gamma-ray tail, i.e., that the observed spectrum is basically the
sum of two thermal Comptonization spectra. While a sum of thermal
Comptonization spectra may indeed match the observations, there is
one potential physical problem with this interpretation. In the
hard state of Cyg X-1, it appears that the total optical depth
of the source is at best $\tau_T \sim 1-2.$ Thus photons are
presumably
free to scatter between the different thermal zones, and 
a uniform bi-Maxwellian electron distribution is not a bad
first approximation to this case. What spectrum does one 
obtain in this case? As shown in Figure~4, while the overall 
spectrum does have an extended high enery tail, the overall spectrum is 
most closely approximated by a {\it single} thermal Comptonization spectrum
with temperature $T_e^{av} \approx (T_e^{lo} +T_e^{hi})/2.$ 
(This is the temperature of  thermal plasma that gives roughly
 the same photon energy change as in the bi-Maxwellian case.)
Adding together a 10 keV and a 100 keV Comptonization spectra
in analogy with the way a multi-color
disk black body is computed will give a very wrong answer in this case. Note
also that a bi-Maxwellian plasma where one component is very hot
has a nasty feature that may rule it out in the case of Galactic
Black Hole Candidates like Cyg X-1. As
we will discuss in more detail below, the fractional change
in the energy of a scattered photon is not small if $kT_e \sim
m_ec^2.$ This means the multiple Compton scattering approximation
begins to break down, and one can see evidence of
the first scattering order -- in Figure~4, a $\sim 0.1-0.3$ keV
{\it deficit} of photons relative to a low-energy extrapolation of 
the 2-10 keV power law.

In conclusion, we remark that if only
one or two orders of scattering contribute significantly to
the emergent spectrum (e.g., in a non-thermal model that is 
not photon-starved), then the shape of the spectrum
is obviously extremely sensitive to the details of the
underlying electron energy distribution. We do not have
space to discuss the time-dependent behavior of hybrid models, 
e.g., low vs. high energy lags and leads, but any such behavior 
will clearly depend on whether the emergent spectrum
is produced in the multiple  or single Compton scattering 
regimes. In a one-zone model, if the spectrum is produced in one scattering
(e.g., in non-thermal models), one expects  no delay between
different photons energies except perhaps for a slight soft lag
due to the finite time it takes for electrons to cool
and respond to changes in injection (which can be much 
shorter than $R/c$ and thus hard to observe). If multiple
Compton scattering is instead important, one expects to see 
behavior similar to that seen in standard thermal Comptonization models,
e.g., hard lags that increase logarithmically with energy.

\section{An Application of a Hybrid Model to Galactic Black Hole
Candidates}

Until now, we have mainly discussed hybrid plasma models in
a theoretical context. Is there any observational evidence
that they may be important? In Active Galactic Nuclei (AGN), 
the situation is still
unclear. Radio-loud AGN probably have X-ray emission that is contaminated
by strongly non-thermal emission from a jet, and 
they will not be considered here. (The jet environment 
in the emission region may 
be very different from that near the black hole.) For radio-quiet AGN,
the composite Seyfert spectra compiled by OSSE (e.g.,
Gondek et al. 1996) favor a spectral cutoff at an energy
$\sim 300-500$ keV. An equally good fit to the composite
spectrum is obtained using either
a purely thermal model or a purely non-thermal model with either steep
power law injection or a low $\gamma_{max}$ (maximum electron injection
energy),
i.e., the energy coverage
and statistics of the composite spectra are not 
sufficient to distinguish between pure thermal,
hybrid, or pure non-thermal models. It is also not clear, however,
how representative the composite spectra are. For example, Matt 
(these proceedings) reports on a BeppoSax Seyfert sample that 
shows considerable variation in the cutoff energies,
from $\sim 70$ keV in NGC 4151 to beyond $\sim 200$ keV in several
objects. The strong break in NGC 4151 favors a purely thermal
or hybrid model (e.g., Zdziarski, Lightman, \& Maciolek-Niedzwiecki
1993), but fits to other individual objects are inconclusive.

\begin{figure}[t]
\centerline{\epsfig{file=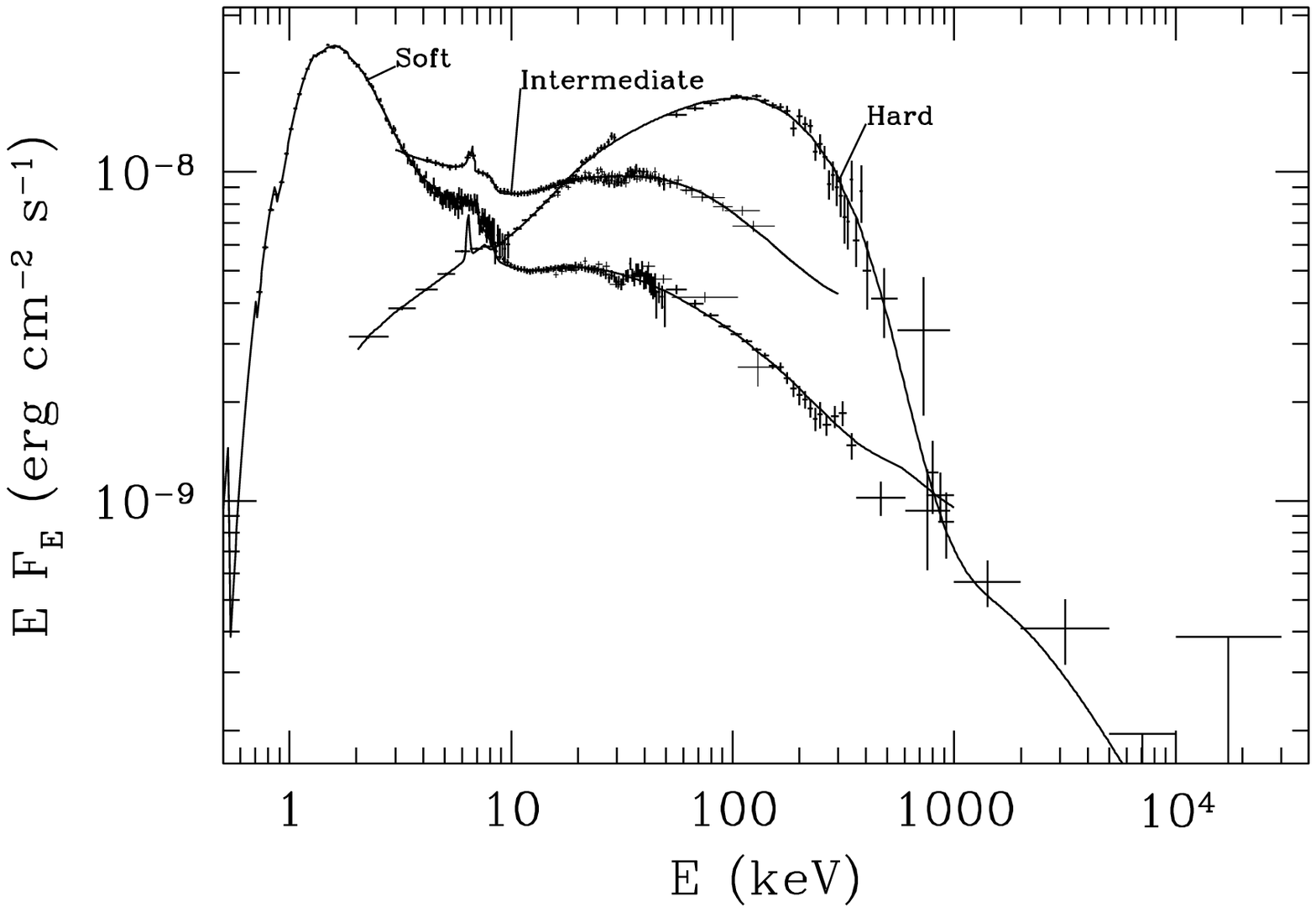,width=13cm,height=8.5cm}}
\caption{Spectral states of Cyg X-1. The hard state spectrum
shown consists of simultaneous {\it Ginga} and OSSE observations
taken on June 6, 1991 (Gierli\'nski et al. 1997) and an overlapping,
longer duration observation taken by Comptel between May 30--June 8, 1991
(McConnell et al. 1994). The intermediate state spectrum  was obtained by RXTE
on May 23, 1996. The soft state spectrum was obtained by ASCA and RXTE
on May 30, 1996 and by OSSE between June 14-25, 1996. The spectral data
have been rebinned for clarity. The {\it solid} curves show the best fit
{\sc eqpair} models to each of the states. See Gierli\'nski et al. (1999)
for more details on the data and model fit parameters.}
\label{fig-5}
\end{figure}

The situation is potentially more interesting for
Galactic Black Hole Candidates. 
Power law emission extending beyond $511$ keV has
definitely been detected by OSSE during the ``soft
state'' of these objects (e.g., Grove et al. 1998,
Gierli\'nski et al. 1999), and
in objects like Cyg X-1, 
we have observed several spectral
state transitions from the ``soft'' to the ``hard'' state and back (e.g., see 
Liang \& Nolan 1984; Cui et al. 1997). Here, I will focus on
Cyg X-1 as a test case since the object always seems to be 
bright, and hence considerable data has been collected on 
it. Figure~5, taken from Gierli\'nski et al. 1999, 
shows a montage of spectra obtained during the soft,
hard, and intermediate (transitional) states of Cyg X-1.
Simultaneous, broad band data of this quality has had 
and will continue to have (as the instruments improve) an important
impact on our understanding of this object.  Until simultaneous
10-100 keV observations were available, for example, it was
impossible to constrain the contribution to the overall
spectrum from a Compton reflection component since the amplitude
of this component depends critically on the hard tail 
($\gtrsim 100$ keV) of the spectrum. Also, it was impossible
to tell how the bolometric luminosity varied with the state
of the source, and it was difficult to constrain the 
parameters of the Comptonizing cloud that was supposed
to be responsible for the spectrum (at least in the hard state).

Using simultaneous
Ginga-OSSE data, however, Gierli\'nski et al. (1997) was able
to show that the standard static disk-corona slab geometry 
was ruled out for Cyg X-1's hard state because: 
(i)  no ``anisotropy break'' was seen, 
(ii) the X-ray spectrum was very hard indicating that the 
source was photon-starved with $l_h/l_s > 1$ 
(in the disk-corona geometry the
soft reprocessed photon luminosity is comparable, i.e.,
$l_h \sim l_s$), and (iii) the solid angle covered
by the reflecting matter substantially less than 
$2\pi,$ the value expected in the disk-corona where
the corona extends uniformly over the disk.
They suggested that a source geometry consistent with
these would be one like that of Shapiro, Lightman, \& Eardley
(1976), with a central, hot source surrounded by a cool,
thin accretion disk (e.g., see Fig. 9). Relying on a composite
Cyg X-1 spectrum made of non-simultaneous data, Dove et al. 
(1997) came to a similar conclusion. 

Hard state data of this type in combination with broad band data on 
the transition to the soft state also led to new suggestions
for the overall accretion disk/corona geometry  in Cyg X-1.
 The data of Zhang et al. 
(1997) showed that the bolometric luminosity in fact 
did not change significantly
($\lesssim$ factor two) during the recent hard-soft state transition.
 Because the luminosity of the blackbody component in the soft
state was comparable to the hard photon luminosity in the hard state, 
and because a thermal Comptonization model for the hard tail
required a low Compton $y$ parameter (a combination of low
temperature and/or optical depth), Poutanen, Krolik, \& Ryde
(1997) proposed that the state transition was simply a change
in the state of the  accretion disk:  the power dissipated in the corona
dropped as the inner edge of the cool disk moved inwards, 
correspondingly increasing the soft photon luminosity. As the cool
region of the disk spread inwards, the fraction of the coronal
emission intercepted by the cool disk increased (e.g., 
as the disk penetrated into the coronal region), causing the
observed increase in the relative amplitude 
of the Compton reflection component. Esin et al. (1998) 
proposed a rather similar model, where the inner radius of the 
cool disk is interpreted as the transition radius between
the Sunyaev-Shakura disk solution and the ADAF solution. While the
Poutanen, Krolik \& Ryde model is more phenomenological and the
discussion in Esin et al. is framed in the more physical context
of ADAFs, both models
are essentially the same, with the key free parameter controlling
the ``state'' of the system being the location of
the inner edge of the cool disk/the disk transition
radius (something not currently well-understood). 
Both models also share the significant shortcoming that neither can 
simultaneously fit the low-energy (1--10 keV) and high-energy
($>300$ keV) data in the soft state. This has not been completely
appreciated and is one of the strongest arguments for the presence
of a hybrid plasma.

The reason both models fail is that they 
rely on purely thermal Comptonization to produce the
observed spectrum.  The X-ray spectrum above $\sim 2$ keV
(e.g., in the RXTE data) appears to be a rather steep 
power law that joins smoothly onto the dominant blackbody
component at $\sim 1$ keV (e.g., see Fig. 5). A Comptonization
model fit to data below $\sim 30$ keV will give something 
like $30-40$ keV as the best fit
electron temperature in the model. At first sight,
a low temperature like this is exactly what one wants
since in the soft state, less energy is being
dissipated in the corona and there are more soft photons
to cool on. Unfortunately, such a low
temperature also predicts an exponential cutoff in the upscattered
spectrum starting at $\sim 3 kT_e \sim 100$ keV. Such a cutoff
is {\it not} seen in the OSSE data for the soft spectra
of Galactic Black Hole Candidates, which in some
cases clearly extend to at least 500 keV and above
(see Grove et al. 1997, 1998). Recently Gierli\'nski
et al. (1999) has taken OSSE data for Cyg X-1 and tried to make as simultaneous
fits as possible to lower energy ASCA and RXTE data. Figure~6
shows an example of joint RXTE-OSSE data that indicates no 
strong break out to $\sim 200$ keV. 

\begin{figure}[t]
\centerline{\epsfig{file=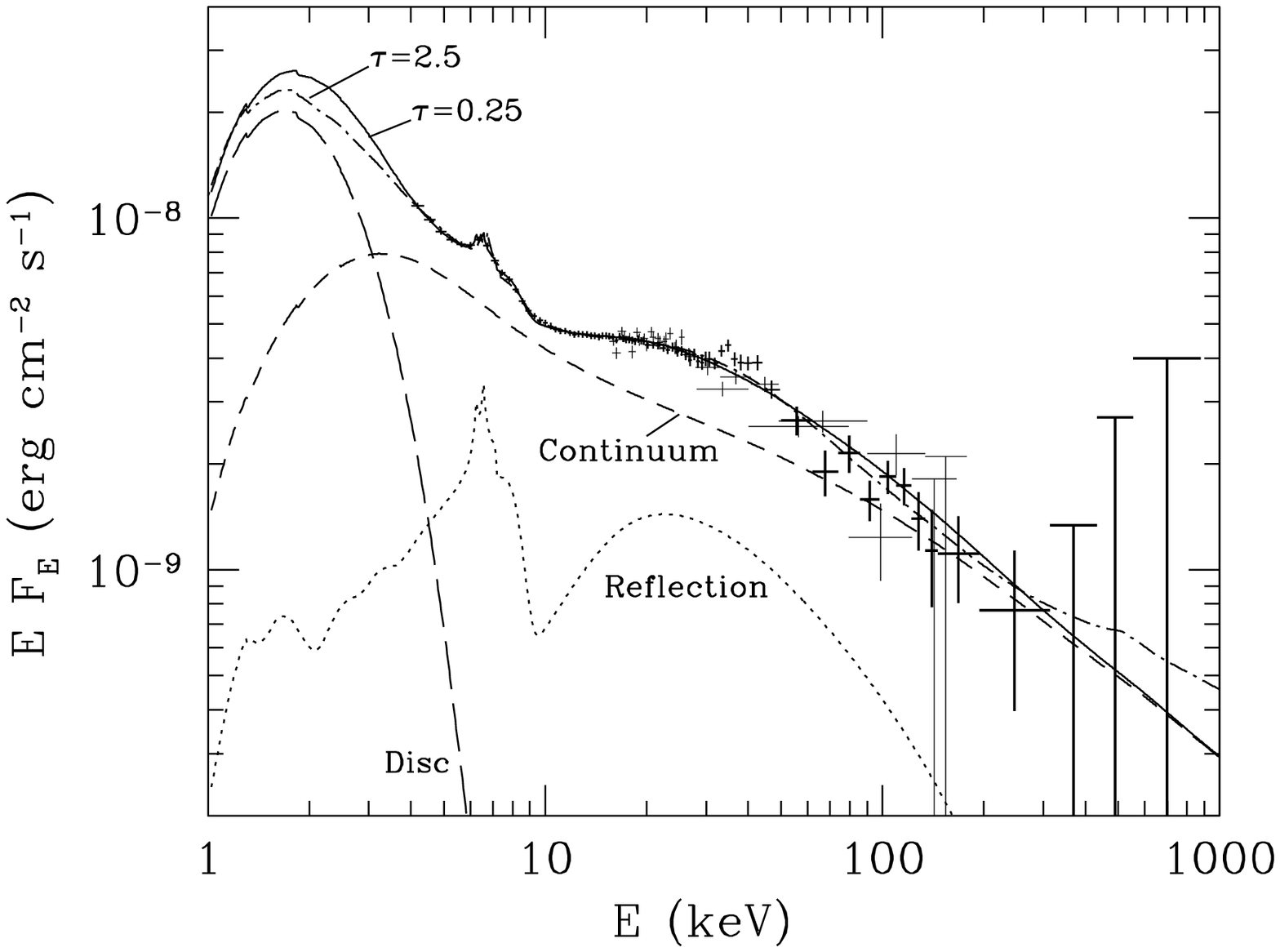,width=13cm,height=8.5cm}}
\caption{Simultaneous RXTE and OSSE observations
of Cyg X-1 in the soft state on June 17, 1996. The curves
show the various model components. The continuum is computed using 
{\sc eqpair}. The parameter $\tau=0.25$ and $2.5$ is the optical depth
of the background electron plasma. Energy is supplied 
{\it only} to non-thermal
electrons ($l_{th}=0$), which are injected with a steep power 
law with number index 
$\Gamma\approx 3$ (see Gierli\'nski et al. 1999 for more details).
Cooled electrons thermalize and share their remaining energy
with the background plasma electrons.}
\label{fig-6}
\end{figure}

In order to produce such
a unbroken spectrum, the temperature of the Comptonizing electrons
must be comparably high. We illustrate this in Figure~7,
where we show the Comptonized spectrum from
electrons with temperature 200 keV. At $\gtrsim 10$ keV,
this spectrum has the right slope to match the soft state
spectrum, and it starts cutting off exactly when the
statistics of most detectors become very poor, i.e., it looks like
an acceptable model.
However, notice the large
photon deficit at $\sim$ 1 keV (briefly mentioned above). The
deficit occurs because the mean fractional energy change 
a blackbody photon undergoes in one scattering is $\Delta \epsilon/
\epsilon \sim 4kT_e/m_ec^2.$ For $kT_e=200$ keV, this exceeds
unity and implies that one can begin to see the shapes of 
the individual Compton scattering orders, particularly the first one.
(Approximating the blackbody soft photon distribution as 
a delta function at energy $\epsilon_{bb},$ there are
few upscattered photons between $\epsilon_{bb}$ and $\sim 2\epsilon_{bb}$
since $\Delta \epsilon$ is so large -- hence we see a deficit.)
Note the good agreement between the {\sc eqpair} result and
the Monte Carlo simulations shown there. 
(When making Comptonization calculations, 
especially using a kinetic code like {\sc eqpair}, one has to be very careful 
not to create spurious features similar to this one by
using an approximate Compton redistribution function that does not
spread scattered photons sufficiently in energy.) Such 
a feature is real and generic to high temperature Comptonization
models -- and is strongly ruled out by data like that 
of Figure~6 at the many
sigma level. (Remember, flux determinations in the keV
range are now good down to the $\sim$ few percent level.)
The discrepancy with purely thermal models is probably even greater
because Gierli\'nski et al. (1999) have averaged together
several days of soft state OSSE data and find that 
any cutoff must be at energies $\gtrsim 800$ keV(!).

\begin{figure}[ht]
\centerline{\epsfig{file=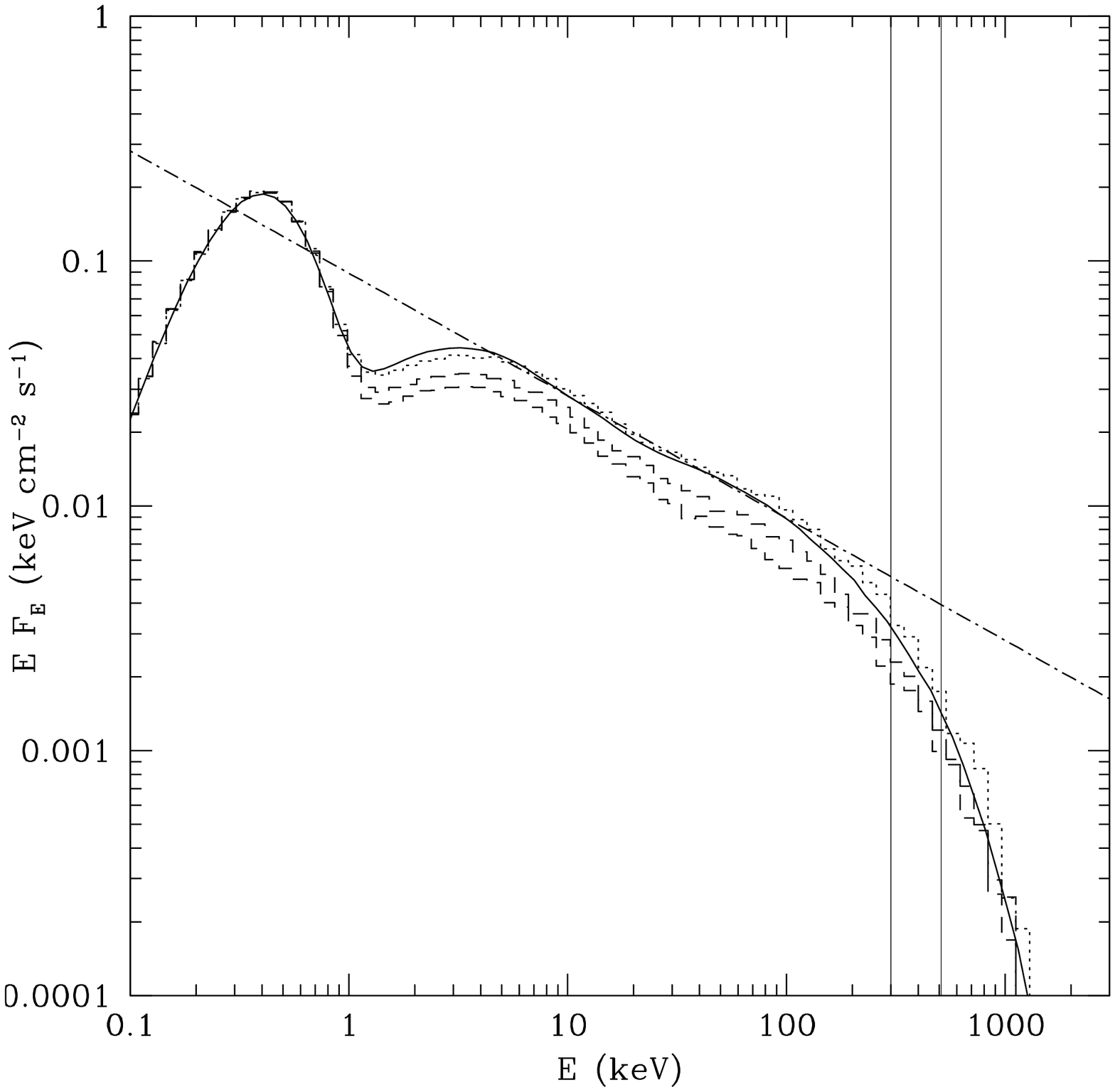,width=11cm,height=8cm}}
\caption{The Comptonization spectra produced by a thermal
plasma with temperature $200$ keV in a spherical source with 
Thomson optical depth $\tau_T=0.1.$ The input soft photon blackbody
spectrum has temperature $kT_{bb} = 0.15$ keV. The {\it solid} curve
shows the result obtained using the {\sc eqpair} Comptonization routine.
From top to bottom, the histograms respectively show results from a
Monte Carlo Comptonization code of M. Gierli\'nski where
photons are injected at the center of the sphere, throughout
the sphere according to the distribution $n(r) \propto
\sin(kr)/r$ of Sunyaev \& Titarchuk (1980), and uniformly throughout
the sphere. The {\it dot-dashed} curve shows a power law with
energy spectral index $\alpha_x=-1.5$ (typical of what is observed
in the Cyg X-1 soft state), and the vertical lines delineate
the $\sim 300-500$ keV region where most (but not all!) 
Galactic Black Hole Candidate observations run out of statistics.}
\label{fig-7}
\end{figure}

If the soft state is not the result of purely thermal Comptonization,
what are the alternatives? First, one might interpret the lack
of a $\sim$ keV deficit as implying the existence of ``excess''
emission. By superposing emission from two spatially distinct
regions (or else one runs into the problems discussed
in the previous section if photons can sample both
regions), one can in principle reproduce
the observed spectrum using a low temperature and
a high temperature region. One then needs to explain, however,
where the extra component comes from and why it has 
the temperature and optical depth it does. Another possibility that has 
recently been revived by Titarchuk and collaborators
in the context of Galactic Black Hole Candidates
is that of bulk Comptonization (e.g., see Blandford \& Payne 1981,
Colpi 1988, Shrader \& Titarchuk 1998, Titarchuk \& Zannias
1998; Psaltis \& Lamb, in these proceedings). The rationale for invoking
bulk Comptonization during the soft state is that the soft
photon density appears to increase dramatically during the soft
state, strongly cooling the coronal electrons
responsible for the hard state emission. If there is quasi-spherical
accretion near the black hole, then cold infalling coronal electrons 
could acquire substantial velocities $v \sim c,$ and Comptonization
using the large bulk inflow velocity of the electrons (as opposed
to their assumed smaller thermal motions) could give rise to 
power law spectra like those observed in the soft state.

While elegant, this interpretation may also have significant problems
fitting data, particularly in the case of Cyg X-1. One of the
key difficulties is again how to produce unbroken power law emission
well beyond 511 keV ($m_e c^2$). 
As a first order estimate of the location of the high
energy break in the bulk Comptonization spectrum, Titarchuk,
Mastichiadis, \& Kylafis (1997) give
$\epsilon
\sim {4 \over \dot m} m_ec^2,$ where $\dot m=\dot M c^2/
L_{Edd}$ is the 
mass accretion rate measured in units of the 
Eddington luminosity and the scattering electrons
are assumed to be radially free-falling. However, to have
a predicted X-ray spectral index in the observed range 
($\alpha_x \sim 1.5-1.8$) apparently
requires a very high mass accretion rate, 
$\dot m \gtrsim 4$ (Titarchuk \& Zannias 1998).
The first order estimate thus predicts a
strong break in the Comptonized spectrum at 
energy $\epsilon \lesssim m_e c^2.$ It is currently still not
clear, though, how good this first order estimate is.
As pointed out in Titarchuk, Mastichiadis, \& Kylafis (1997),
the second order terms in their equations tend to push their
cutoff to higher energies. However, their treatment and essentially
all other treatments until now have worked
in the diffusion approximation and used the Thomson cross-section
with a down-scattering correction 
instead of the full Klein-Nishina cross-section.
As in the case of standard thermal Comptonization, when electron
and photon energies exceed $\sim 0.1 m_e c^2,$ the results obtained
with these approximations become suspect. In particular, 
in the Klein-Nishina limit, Compton scatterings with the more
energetic (higher velocity) electrons are reduced, 
and scattered photons tend to 
keep traveling along their original direction, e.g., into
the black hole. In addition, even moderately relativistic electron
velocities ($v/c\gtrsim 0.3 $) will strongly collimate incoming
radiation along the inflow direction. 
Gravitational redshift effects when infall velocities are $v/c \sim 0.9$  
might further lower the break energy. 
I also note that the soft state temperature
Esin et al. find for the central ADAF region is still $\sim
40$ keV, i.e., it is not that low at all. Standard
thermal Comptonization effects may still dominate over bulk Comptonization
ones, and at the very least, may  cause a significant deviation from a 
power law spectrum. Although a better calculation is required for
a definitive answer, right now it seems difficult to produce
a spectrum beyond 511 keV ($m_e c^2$) via bulk
Comptonization.   How serious a problem is the low predicted break
energy? In most Galactic Black Hole Candidates, the typical data 
above $\sim 200$ keV are not good enough to say anything. In Cyg X-1, however,
we have a clear indication from OSSE data that the spectrum
continues unbroken to at least $\sim$ 800 keV. Although the Cyg X-1
data is not quite as certain at energies higher than this 
due to possible source confusion, as noted above
there is a strong indication that the spectrum 
continues to  several MeV. If correct, this would seem to strongly rule
out the bulk Comptonization hypothesis, for Cyg X-1 at least.
Break energy aside, bulk Comptonization models have one other
serious problem fitting Cyg X-1. Gierli\'nski et al. (1999) finds that
the ASCA/RXTE-OSSE data above $\sim 10$ keV is {\it not}
well-fit by a power law and requires excess emission in 
the $\sim 10-30$ keV range. Together with the presence of an
apparent iron line and edge, this strongly suggests the 
presence of reflection with a covering factor $\Omega/
2\pi \sim 0.7,$ i.e., half the hard X-ray flux hits a cool disk.
Such a covering factor is natural in a corona-over-disk geometry, 
but it is not obvious how to arrange this in the bulk Comptonization
scenario.

\begin{figure}[t]
\centerline{\epsfig{file=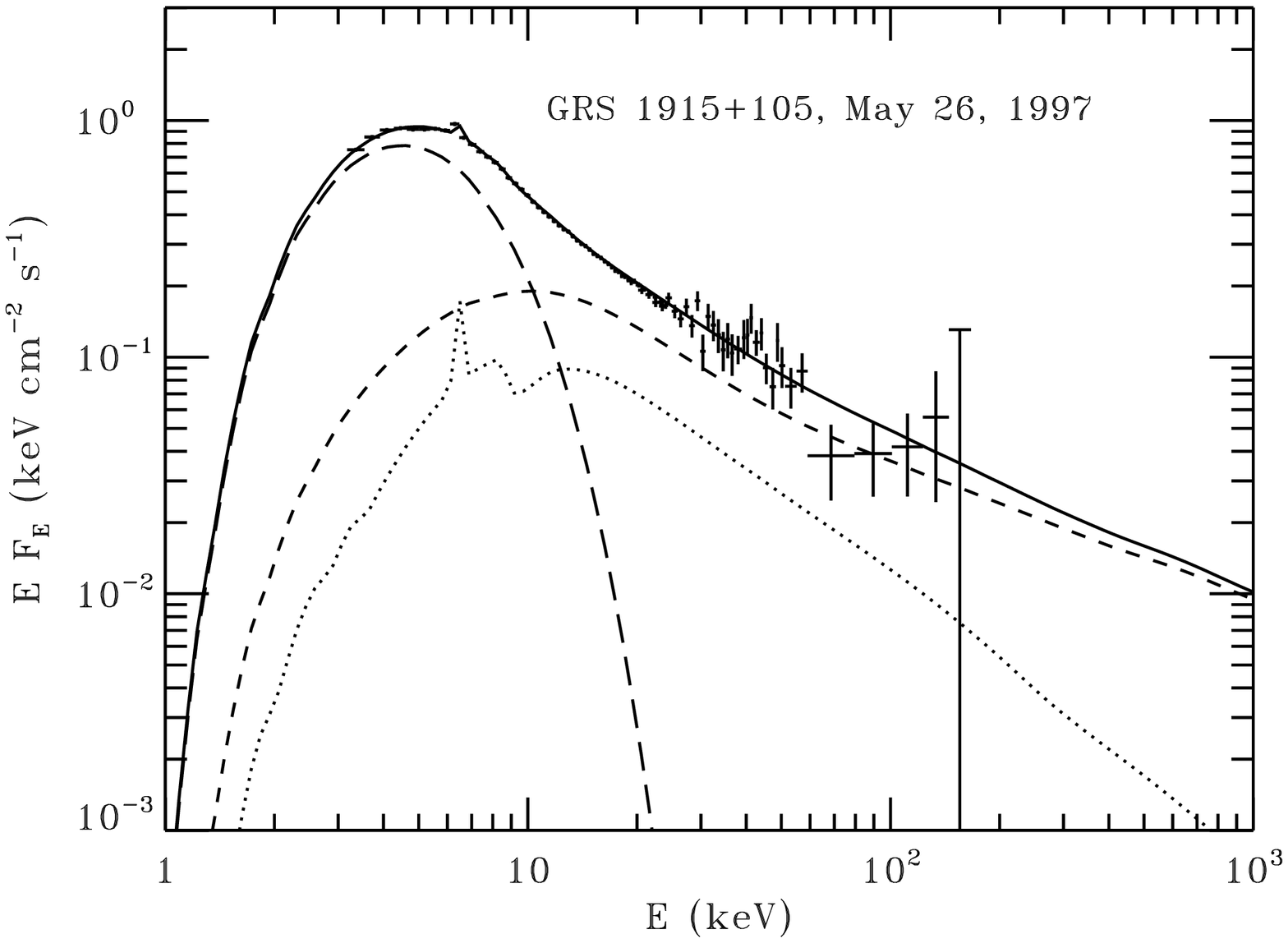,width=12cm,height=9.5cm}}
\caption{A fit to public RXTE data on GRS 1915+105 using
the hybrid model, {\sc eqpair}. The main fit parameters are
$l_s=20,$ $l_h/l_s=0.2,$ and $l_{nth}/l_h=0.6.$ The background
plasma electron optical depth was $\tau_p=0.2,$ and 
non-thermal electrons were injected from $\gamma_{min}=1.3$
to $\gamma_{max}=1000$ with number index $\Gamma=3.$ The reflecting
material subtended a solid angle $\Omega/2\pi=1.1$ as seen
from the source. The soft
photon input was a multi-color disk blackbody ({\sc diskbb}) with
temperature $kT_{mdbb}=1.56$ keV. The reduced $\chi^2$ for the
fit was  1.02 for 208 degrees of freedom. The various model
components used and shown are the same as in Figure~6. }
\label{fig-8}
\end{figure}

As discussed in Poutanen \& Coppi (1998), an easy way to avoid
all the problems mentioned is simply to allow the Comptonizing
plasma to be a hybrid one, with a non-thermal tail.
This introduces an extra parameter, the ratio of thermal heating
to non-thermal acceleration power, but allows disk transition scenarios
like those of Poutanen et al. and Esin et al. to go through
largely unchanged.  As shown by the solid curves in Figures~5 and 6,
the hybrid model ({\sc eqpair}) including Compton reflection and 
relativistic line smearing can fit the broad band data extremely
well ($\chi^2$ per d.o.f $\sim 1$) in {\it all} three states of Cyg X-1,  
including
the ``intermediate'' transitional one. To see how far we could
push the hybrid model, we also applied it to public RXTE data on GRS 1915+105
and also were able to obtain a good fit, e.g., see Figure~8 (although
that data only extends to $\sim 150$ keV).
 We are not aware of another
type of model that can currently do this. This is slightly  surprising given
the quality of the ASCA/RXTE data in the 1-20 keV range and may 
indicate that the crude assumptions made in the {\sc eqpair} model
(e.g., that the source is homogeneous, isotropic, and static)
are true to first order.
A detailed discussion of the model fits to Cyg X-1 and 
their physical implications (e.g., constraints on the
importance of electron-positron pairs) can be found
in Gierli\'nski et al. (1999).

I conclude by showing results a simple phenomenological model
which gives roughly the right fit parameters for the states
(see Poutanen \& Coppi 1998 for more details). The source geometry
we envision is along the lines of that shown 
in Figure~9.
The total power supplied to the disk and the $\;$ 
hot $\;$ coronal  $\;$ region 
\newpage 

\begin{figure}[ht]
\centerline{\epsfig{file=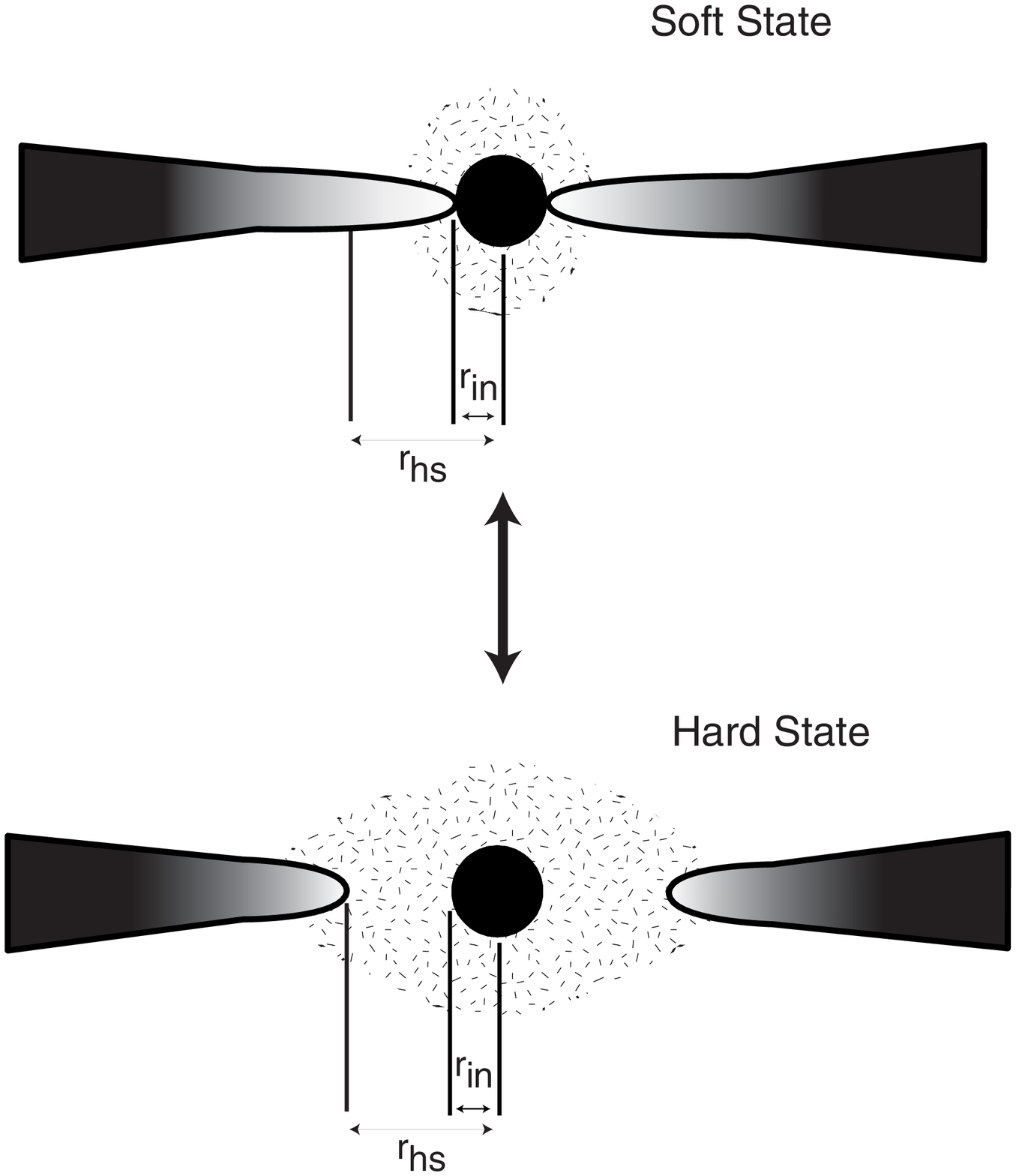,width=12cm,height=9.5cm}}
\caption{\small 
A schematic diagram showing the source geometries
envisioned in the Poutanen \& Coppi (1998) model for 
the Cyg X-1 transition. In the hard state, the inner part of 
the disk puffs up and acts as a hot, Comptonizing corona.
(Or alternatively, see text, the hard state could simply
represent a significant increase in the dissipation rate of
accretion power in a corona above.)
The hot inner disk/corona is surrounded by a cool Shakura-Sunyaev
disk which is the source of the soft seed photons which
enter the inner Comptonizing region. Non-thermal electrons
are also injected into the hot inner region by an 
unspecified acceleration mechanism. Some of the 
escaping Comptonized photons are intercepted by the disk
and Compton reflected back to the observer, producing the
fluorescent iron line, the iron edge, and the reflection
``hump'' at $\sim 10-30$ keV. In the transition to the soft 
state, the edge of the cool disk moves inwards, perhaps
very close to the last marginally stable orbit, and penetrates
into the inner coronal region. Because the hot region of the disk is
now mostly  gone,
the thermal power effectively supplied to the corona is very small.
Only the non-thermal accelerator continues to make a significant
contribution to the corona's power. 
}
\label{fig-9}
\end{figure}

\noindent 
remains roughly constant during the
state transition, i.e.,  $L_{rad}=L_s + L_h$
is constant. As in the Poutanen et al.  and Esin et al. models,
we assume a transition radius $r_{tr}$ which marks the 
boundary between a cool outer disk and a hotter inner disk/corona.
We also assume that the sum of soft luminosity from the disk, 
$L_s\propto 1/r_{tr}$, and 
the thermal dissipation rate in the corona, $L_{th}\propto 1-1/r_{tr}$, 
remains approximately constant during transitions.
In addition to thermal dissipation in the corona,
we will assume a central source of non-thermal electrons with
acceleration luminosity, $L_{nth}$. (Note that if the source turns
out to have a relatively low compactness, the non-thermal 
acceleration could also occur in a more extended 
region that does not change size. One will obtain quite
similar spectra to those shown here.) 
For simplicity, we take 
$L_{nth}$ to be constant (its value is not that well-constrained
in the hard state). 
In the soft state, essentially all the power goes into accelerating
non-thermal electrons. The density and temperature of the thermalized
electrons/pairs responsible for
the excess emission at $\sim$ a few keV are determined self-consistently. 
The results of the simulations are shown in Figure~10.

\begin{figure}[h]
\centerline{\epsfig{figure=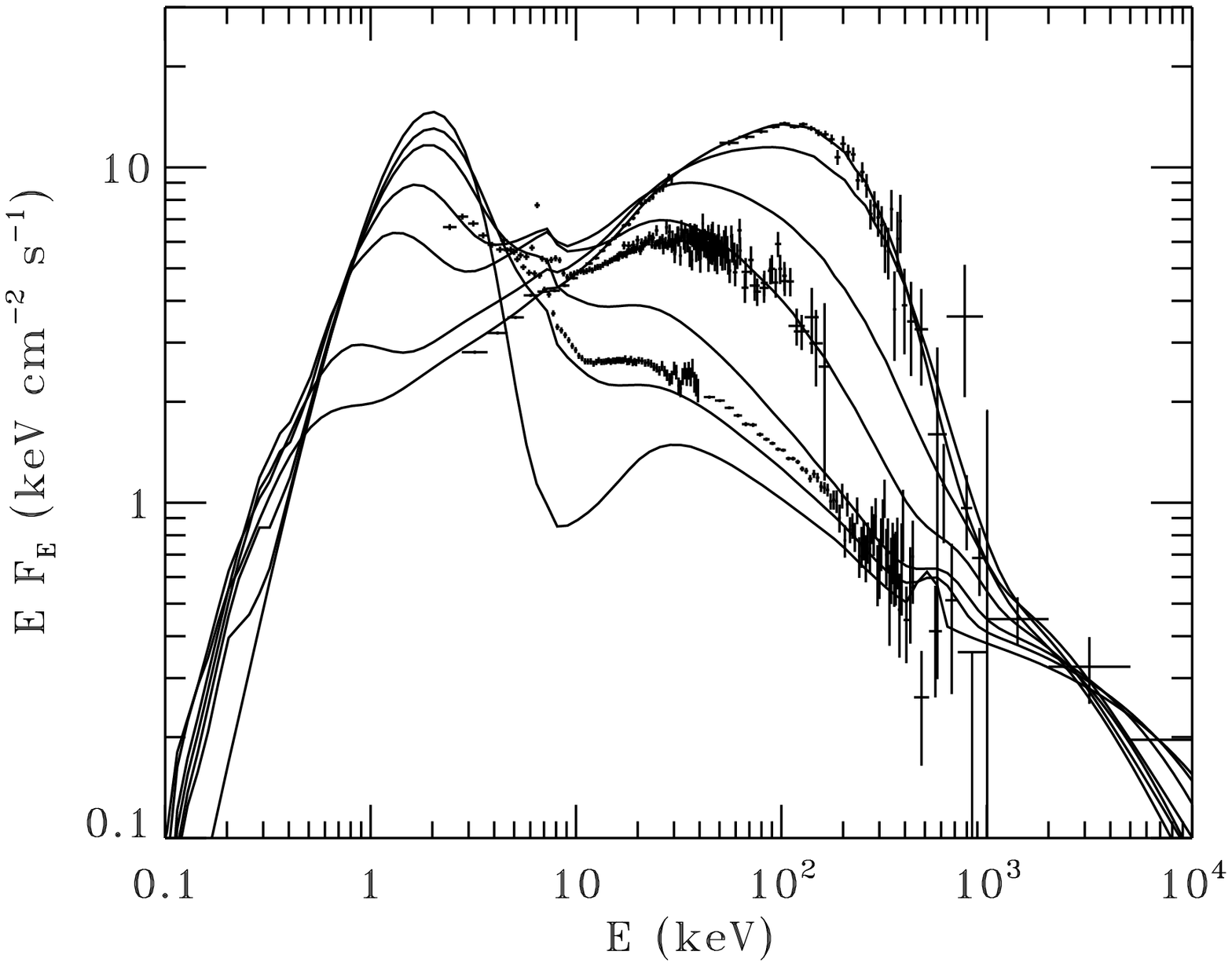,width=13cm,height=8.5cm}}
\caption{
Simulations of the Cyg X-1 spectral transition using the hybrid pair model,
{\sc eqpair}. The starting point is the best fit to the hard state data.
Keeping $L_{rad}=L_s+L_h$ constant and using some simple scaling
laws  for $L_s, L_{th}, L_{nth}$  and $\tau_p$ (see Poutanen \& Coppi 1998), 
we obtain the sequence of spectra covering  the transition between
the hard and soft states as a function of the transition radius, 
$r_{tr}$. (Large $r_{tr}$ gives the hard state; low $r_{tr}$ 
gives the soft state.) 
}
\label{fig-10}
\end{figure}

Although we have interpreted the state transition as a
change in the location of the inner edge of the cool disk, note that
this interpretation is in fact not unique. The arguments used in
Gierli\'nski et al. 1997 apply {\it only} to models where the corona
is static, i.e., there is no bulk motion of coronal electrons.
As discussed in Beloborodov (1999), this is rather unlikely
given the strong radiation fields likely to be present and also
given the highly dynamic, flaring nature of the emission. 
(The possibility of winds arising from the surface of disks
has also been discussed for some time,  e.g., see Narayan \& Yi
1995, Blandford \& Begelman 1998 in the context of ADAF-type
solutions; and Woods et al. 1996 for a numerical calculation of X-ray
heated winds and coronae from accretion disks.) When the electrons
acquire even relatively small outflow velocities ($v/c \gtrsim 0.2$),
this is enough to make their upscattered radiation significantly 
anisotropic (see Beloborodov 1999). Even if a cold disk extends all the
way in to the last marginally stable orbit and is directly
under the corona (e.g., the slab disk-corona geometry again),
this anisotropy means that fewer hard photons reach the disk, i.e., the 
inferred covering fraction ($\Omega/2\pi$) of the Compton reflecting  
material will be low. If most of the accretion power near
the black hole is dissipated in the corona above the 
disk rather than inside the disk, then this also means the disk underneath
the corona will be very cold and will emit few soft photons,
i.e., the coronal plasma will be photon-starved. (This is
in contrast to the static corona of Haardt \& Maraschi 1993
where half the hard X-ray flux hits the corona and is reprocessed,
implying a soft photon luminosity comparable to the hard photon
luminosity.) In other words, if there is bulk motion in the 
corona, then there is no need to have a transition from
a cool to a hot disk. Rather, we might have a transition from a
cool to a {\it cooler} disk. In this case, the transition radius
in our model is to be interpreted as the boundary between
the region where most dissipation occurs inside the disk
and the region where (for some reason) most dissipation occurs 
outside the disk, in the corona. As long as the coronal outflow
velocities are not too large, the results are quite
similar to those of the central hot sphere-disk geometry
of Figure~9.

\section{Summary}

Most attempts at modeling the emission from accreting black hole
systems have typically assumed the underlying 
particle energy distribution is either a Maxwellian (``thermal'')
or a power law (``non-thermal''). While such an assumption may
be convenient analytically, it is no longer required given
the advent of powerful computers, and more importantly, it
is does not appear to be well-justified. There are several
examples in Nature, e.g., the phenomenon of solar flares, where
it is clear the underlying distribution is neither a Maxwellian
nor a power law, but rather a quasi-Maxwellian at low energies
with a high-energy approximately power-law tail. This is exactly
the type of particle energy distribution that is often predicted
by theoretical particle heating/acceleration models. (The
acceleration process typically kicks only a few particles in the high
energy tail of the particle distribution to much higher energies.) 
Re-examining
the process of electron thermalization under the physical conditions
likely to be found near a black hole, we find that the likely
thermalization time scales are likely to be quite long
-- unless some (unknown) collective plasma process is
more effective than two-body Coulomb collisions at
exchanging energy between electrons. In particular, because
the radiation field is likely to be so intense near an
accreting black hole, the Coulomb relaxation time for even
moderately relativistic electrons may be much longer than the relevant
cooling times. Depending on the exact plasma parameters, they may
also be longer than the characteristic source variability time.
Thus, the old arguments made against thermal models still 
stand. These arguments were largely brushed aside and forgotten
when it became clear many of the classical non-thermal sources
like  NGC 4151 (an AGN that was supposed to have strong
MeV emission) in fact showed strong cutoffs at $\sim 100$ keV
energies and had spectra that could be successfully fit using
purely thermal Comptonization models. 

Now that we are seeing
hints that the emission in Galactic Black Holes Candidates
may indeed extend to much higher energies
(albeit at much lower levels than previously thought), ``hybrid'' 
models involving both thermal particle heating and non-thermal 
particle acceleration are beginning to creep back.
One
of the strongest cases for the existence of a non-thermal
particle distribution near black holes is the ``soft'
spectral state of Galactic Black Hole Candidates. In this state,
one sees very strong, quasi-blackbody emission at $\sim$ 1 keV,
with a steep $\sim$ power law tail extending to at least $\sim$ 511 
($m_e c^2$) in several objects.  Particularly in the case of 
Cyg X-1, such emission is very hard to explain either via 
pure thermal
Comptonization or bulk Comptonization in the accretion flow.
This has not been completely appreciated. Data from future missions
with improved sensitivity in the $\sim 500$ keV - $1$ MeV range
(e.g., INTEGRAL and Astro-E) should be conclusive.
If we relax the assumption that the energy distribution
of the Comptonizing electrons is a strict Maxwellian, then
many problems go away. Using a newly developed, self-consistent hybrid plasma
code, we show that the spectrum in the soft state (as well as in the other 
spectral states) can  easily be modeled. With the proviso that
there is always some small amount  of non-thermal acceleration going on
(compared to the total source luminosity), models
that explain the Cyg X-1 state transitions in terms of a moving
transition radius between cold and hot disk phases appear to
work fairly well.  (As Cui points out in these proceedings, however,
these models only attempt to explain {\it time-averaged} spectra.
What such a spectrum and the deductions one makes from such a
spectrum have to do with reality is not yet clear,  
particularly if this time-averaged emission is the superposition
of many individual flare events, e.g., as discussed here by 
Poutanen \& Fabian as well as Mineshige \& Negoro.)

If we follow theoretical prejudice and assume that  particle
energy distributions are indeed not completely thermal, then
we must explain why so many black hole sources still manage to look
so thermal. Clearly, one part of the answer must be that
the efficiency with which power is channeled into
relativistic electrons (Lorentz factors $\gtrsim 10$) is
relatively low. Why this is so, depends
on the unknown details of the 
acceleration mechanism. Another part of the answer, however, may
have to do with  Comptonization and pair plasma physics. If a 
source photon 
gains most of its energy in a single scattering event before
escaping, then clearly the emergent radiation spectrum depends
critically on the underlying scattering electron energy spectrum.
However, if multiple Compton scattering is important, i.e., a photon
gains its energy in several small steps before it escapes, then
the exact details of the energy distribution turn out not to matter.
To first order, if the scattering electrons have the same mean
energy, be they thermal or non-thermal, then they tend to produce
the same mean energy change in a scattered photon, which results in 
a spectrum with the same shape. (One can imagine doing a 
Fokker-Planck expansion of the relevant equations.)  The details
of the distribution typically only 
matter at the high-energy and low-energy tails of the output
spectrum. The types of sources where multiple Compton is most
important are those that are relatively photon-starved,
i.e., where the power supplied to electrons is much larger
than the power initially supplied to the low energy target
photons. In this case, as long as the bulk of the
heated/accelerated electrons do not have Lorentz factors
$\gtrsim 2-3,$ it is largely irrelevant whether the
electrons thermalize or not before they cool. (Note that
if the source is very optically thick to gamma-ray pair production,
and many generations of pair cascading are important, then the 
condition that most electrons have low Lorentz factors is
automatically guaranteed -- even if the accelerated electrons which
initiate the cascading have high initial energies.) 
To model Cyg X-1 completely,
for example, we may need a non-thermal power law source
of energetic electrons. To match the data, however, the 
non-thermal acceleration must also  produce a very steep power
law in energy, i.e., most of the power resides in the lowest
energy electrons. If we add such a particle distribution to, say, a 
hot background thermal plasma distribution, and then make the 
source photon starved, we will see virtually no difference in 
the final spectrum, except perhaps at the highest
energies $\gtrsim 200$ keV. (For connoisseurs of 
pair plasmas, though, note that differences in the high
energy photon spectrum can mean big differences in the pair
balance and the pair thermostat.) It is probably no accident, then,
that the photon-starved hard state of Galactic black holes
looks so thermal, while the photon-rich soft state does not.
In sum, whether we realize it or not, hybrid plasmas may be all
around us!
\section{Acknowledgments} I thank Juri Poutanen, Andrei Beloborodov,
and Roland Svensson for organizing an excellent conference, for 
many useful discussions and  hospitality in past, and most of
all for their patience and help in getting this contribution out.  
I would also like to thank Marek Gierlinski and Andrzej Zdziarski for
helpful discussions and allowing me to use their Monte Carlo 
Comptonization code. This work was supported in part by NASA grants 
NAG5-6691 and NAG5-7409.


\begin{references}

\reference Begelman, M.~C. \& Chiueh, T. 1988, \apj, 332, 872

\reference Beloborodov, A.~M. 1999, \apjlett, 510, L123

\reference Benka, S. \& Holman, G. 1994, \apj, 435, 469

\reference Benz, A.~O. \& Krucker, S. 1999, \astap, 341, 286

\reference Blandford, R.~D. \& Begelman, M.~C. 1998,  
\mnras, submitted (astro-ph/9809083)

\reference Blandford, R.~D. \& Eichler, D. 1987, Phys. Rep., 154, 1

\reference Blandford, R.~D. \& Payne, D.~G. 1981, \mnras, 194, 1041

\reference Colpi, M. 1988, \apj, 326, 223

\reference Coppi, P.~S. 1992, \mnras, 258, 657

\reference Coppi, P.~S. \& Blandford, R.~D. 1990, \mnras, 245, 453

\reference Coppi, P.~S., Madejski, G.~M., \& Zdziarski, A.~A. 1999,
in preparation

\reference Crosby, N., Vilmer, N., Lund., N., \& Sunyaev,
 R. 1998, \astap, 334, 299

\reference Cui, W. et al. 1997, \apj, 484, 383

\reference Dermer, C.~D. \& Liang, E. 1989, \apj, 339, 512

\reference Done, C. \& Fabian, A.~C. 1989, \mnras, 240, 81

\reference Dove, J.~B., Wilms, J., Maisack, M., \& Begelman, M.~C. 1997,
\apj, 487, 759

\reference Esin, A.~A. et al. 1998, \apj, 505, 854

\reference Galeev, A.~A., Rosner, R., \& Vaiana, G.~S.  1979, \apj,  229, 318

\reference Ghisellini, G., Guilbert, P., \& Svensson, R. 1988, \apjlett,
335, L5

\reference Ghisellini, G. \& Haardt, F. 1994, \apjlett, 429, L53

\reference Ghisellini, G., Haardt, F., \& Fabian, A.~C. 1993, \mnras,
263, L9

\reference Ghisellini, G., Haardt, F., \& Svensson, R. 1998,
\mnras, 297, 348

\reference Gierli\'nski, M. et al. 1997, \mnras, 288, 958

\reference Gierli\'nski, M. et al. 1999, \mnras, in press

\reference Gondek, D. et al. 1996, MNRAS, 282, 646

\reference Grove, J.~E., Kroeger, R.~A., \& Strickman, M.~S.,
1997, 
     in The Transparent Universe, Proc. 2nd INTEGRAL workshop,
     ESA SP-382, 197 

\reference Grove, J.~E. et al. 1998, \apj, 500, 899 

\reference Guilbert, P.~W., Fabian, A.~C., \& Rees, M.~J. 1983,
\mnras,  205, 593

\reference Haardt, F. 1993, \apj, 413, 680

\reference Haardt, F. \& Maraschi, L. 1993, 413, 507

\reference Li, H., Kusunose, M., \& Liang, E.~P. 1996, \apjlett,
460, L29

\reference Liang, E.~P. 1991, \apj, 367, 470

\reference Liang, E.~P.  \& Nolan, P.~L. 1984, \ssr, 38, 353

\reference Ling, J.~C. et al. 1997, \apj, 484, 375

\reference McConnell, M.~L. et al. 1994, \apj, 424, 933

\reference Maciolek-Niedzwiecki, A., Zdziarski, A.~A., \&
Coppi, P.~S. 1995, MNRAS, 276, 273

\reference Mahadevan, R. \& Quataert, E. 1997, \apj, 490, 605 

\reference Mitsuda, K. et al. 1984, \pasj, 36, 741

\reference Moskalenko, I.~V., Collmar, W., \& Sch\"onfelder, V. 1998, 
\apj, 502, 428

\reference Narayan, R. \& Yi, I. 1995, \apj, 452, 710

\reference Poutanen, J., Krolik, J.~H., \& Ryde, F. 1997, \mnras, 292, L21

\reference Poutanen, J. \& Coppi, P.~S. 1998, Physica Scripta, T77, 57
(astro-ph/9711316)

\reference Rybicki, G. \& Lightman, A.~P. 1979, Radiative Processes in
Astrophysics, New York: John Wiley \& Sons  

\reference Shakura, N.~I. \& Sunyaev, R.~A. 1973, \astap, 24, 337

\reference Shapiro, S., Lightman, A.~P., \& Eardley, D.~M. 1976, \apj,
204, 187

\reference Shrader, C. \& Titarchuk, L. 1998, \apjlett, 499, L31

\reference Stern, B. E., Poutanen, J., Svensson, R., Sikora, M., \& Begelman, M. C. 1995, \apjlett, 449, L13

\reference Sunyaev, R.~A. \& Titarchuk, L.~G. 1980, \astap, 86, 121

\reference Svensson, R. 1984, MNRAS, 209, 175

\reference Svensson, R. 1987, MNRAS, 227, 403

\reference Tajima, T., \& Shibata, K. 1997, Plasma Astrophysics 
(Frontiers in Physics, 98), New York: Perseus 

\reference Titarchuk, L., Mastichiadis, A., \& Kylafis, N.~D. 1997,
\aaps, 120, C171

\reference Titarchuk, L. \& Zannias, T. 1998, \apj, 493, 863

\reference Woods, D. et al. 1996, \apj, 461, 767

\reference Zdziarski, A.~A. \& Coppi, P.~S. 1991, \apj, 376, 480

\reference Zdziarski, A.~A., Coppi, P.~S., \& Lamb, D.~Q. 1990, \apj,
357, 149

\reference Zdziarski, A.~A., Lightman, A.~P.,  \& Maciolek-Niedzwiecki,
A. 1993, \apjlett, 414, L93

\reference Zhang, S.~N. et al. 1997, \apj, 477, L95

\end{references}
\end{document}